\documentclass[12pt]{article}
\usepackage{amsfonts}
\usepackage{dsfont}
\usepackage{amsmath}
\usepackage{amssymb}
\usepackage{amsthm}
\usepackage{geometry}
\usepackage{indentfirst}
\usepackage{graphicx} 
\usepackage{natbib} \bibpunct{(}{)}{;}{a}{,}{,} 
\usepackage[colorlinks=true,urlcolor=blue,citecolor=blue,linkcolor=blue,bookmarks=true]{hyperref}
\usepackage{appendix}
\usepackage{mathrsfs}
\usepackage{enumerate}
\usepackage{tikz}
\usetikzlibrary{matrix, positioning}
\usetikzlibrary{intersections}
\usepackage{subfigure}
\usepackage{caption}
\usepackage{subcaption}
\usepackage[compact]{titlesec}

\newtheorem{thm}{Theorem}[section]

\newtheorem{rmk}{Remark}
\newtheorem{ass}{Assumption}

\newtheorem{Def}{Definition}[section]
\allowdisplaybreaks 
\numberwithin{equation}{section}

\def\qed{ \hfill \vrule height 6pt width 6pt depth 0pt\newline}

\theoremstyle{definition}

\newcommand{\E}{\mathbb{E}}

\newcommand{\R}{\mathbb{R}}

\newcommand{\dif}{\mathrm{d}}

\DeclareMathOperator*{\esssup}{esssup}

\title{Self-protection and insurance demand with convex premium principles}

\date{\today}

\author{Qiqi Li\thanks{Department of Mathematics, Southern University of Science and Technology, Shenzhen 518055, Guangdong Province, P.R. China. Email: \url{12332868@mail.sustech.edu.cn}.} 
\and 
Wei Wang\thanks{Research Institute for Interdisciplinary Sciences, School of Information Management and Engineering, Shanghai University of Finance and Economics, Shanghai 200433, China. Email: \url{wangwei1@sufe.edu.cn}.} 
\and 
Yiying Zhang\thanks{The corresponding author. Department of Mathematics, Southern University of Science and Technology, Shenzhen 518055, Guangdong Province, P.R. China. Email: \url{zhangyy3@sustech.edu.cn.}}$^,$}

\begin{document}

\maketitle

\begin{abstract}
In economic analysis, rational decision-makers often take actions to reduce their risk exposure. 
These actions include purchasing market insurance and implementing prevention measures to modify the shape of the loss distribution. 
Under the assumption that the insureds' actions are fully observed by the insurer, this paper investigates the interaction between self-protection and insurance demand when insurance premiums are determined by convex premium principles within the framework of distortion risk measures. 
Specifically, the insured selects an optimal proportional insurance share and prevention effort to minimize the risk measure of their end-of-period exposure. 
We explicitly characterize the optimal combination of prevention effort and insurance demand in a self-protection model when the insured adopts tail value-at-risk or a subclass with strictly concave distortion functions. 
Additionally, we conduct comparative static analyses to illustrate our main findings under various premium structures, risk aversion levels, and loss distributions. Our results indicate that market insurance and self-protection are complementary, supporting classical insights from the literature regarding corner insurance policies (i.e., null and full insurance) in the absence of ex ante moral hazard. 
Finally, we consider the effects of moral hazard on the interaction between self-protection and insurance demand. 
Our findings show that ex ante moral hazard shifts the complementary effect into substitution effect. 

    
\end{abstract}

\section{Introduction}
\label{sec:intro}

In economic analysis, it is well-established that rational decision-markers commonly take actions to reduce their risk exposure. 
In actuarial science, insured individuals typically purchase insurance products to manage and transfer their risk. 
Since the pioneering works of \cite{borch1960attempt} and \cite{arrow1963uncertainty}, extensive research has explored the identification of optimal insurance structures, such as deductibles and co-insurance, across various perspectives and frameworks (see, for example, \cite{cui2013optimal,chi2013optimal,xu2013optimal,zhuang2016marginal,xu2019optimal,cheung2019budget,zheng2020optimal,chiwei,birghila,jin2024optimal}). 
However, empirical findings and practitioner insights increasingly suggest that insurance contracts alone often fail to provide  comprehensive risk mitigation, especially for certain uninsurable risks or emerging risks where market insurance is unavailable, see, for instance, \cite{vollaard2011does,pannequin2020insurance,awondo2023estimating} and references therein.

Given these limitations in the insurance market and practical insights, rational decision-makers are increasingly combining insurance with prevention measures in risk management. 
This approach not only reduces premiums and potential losses but also aligns with insurers' interests, as insurers often encourage prevention actions to lower claim risks. 
For instance, in health insurance, many companies offer incentive programs that reward prevention efforts by providing premium discounts or reimbursing health-related expenses, such as vaccinations and post-exposure prophylaxis. 
Consequently, examining the interaction between market insurance and prevention efforts is crucial for a more comprehensive understanding of risk management strategies.

The foundational work of \cite{ehrlich1972market} pioneers the study of the interaction between market insurance and prevention efforts, distinguishing two types of prevention: 
\emph{self-insurance} and \emph{self-protection}. 
Self-insurance refers to the efforts that reduce the size of losses, while self-protection refers to efforts that lower the frequency of losses. 
By considering a utility-maximizing decision-maker in a two-state model, they concludes that self-insurance and market insurance are substitutable; thus, an increase in premium loading leads to higher self-insurance efforts but lower market insurance. 
Counterintuitively, they also found that self-protection and market insurance are complementary, meaning that an increase in premium loading raises both prevention efforts and insurance demand. 
Since then, numerous studies in insurance economics have revisited and extended these insights to examine the impact of other variables. 
For example, \cite{konrad1993self} examines self-insurance and self-protection with rank-dependent expected utility preferences and demonstrates that self-insurance effort increases with greater risk aversion, while results for self-protection are less clear. 
\cite{courbage2001self} revisits the findings of \cite{ehrlich1972market} under Yaari's dual theory of choice, which incorporates probability distortion, and justifies that self-insurance and market insurance are substitutes, while self-protection and market insurance are complementary. 
\cite{eeckhoudt2005impact} explore how the prudence coefficient influences optimal prevention strategies, while \cite{dionne2011impact} study the link between absolute prudence and self-protection activities. 
\cite{courbage2017optimal} extend the work of \cite{ehrlich1972market} to include multiple risks. 
\cite{pannequin2020insurance} conduct a laboratory experiment that verifies the substitution effect between self-insurance and market insurance but observes that the substitution effect is less pronounced than what the theory implies. 
The literature mentioned above is far from exhaustive, we refer interested readers to \cite{courbage2013prevention}, \cite{bleichrodt2022prevention} and \cite{peter2024economics} for a more comprehensive review of this topic.

Motivated by the recent work of \cite{bensalem2020prevention} and \cite{seog2024moral}, we revisit the optimal insurance demand and prevention efforts for self-protection under distortion risk measures (DRMs), where insurance is of the proportional form and priced using the convex premium principle used in \cite{cao2024optimal} and \cite{ghossoub2025bowley} with 
the insured's efforts being fully observable or unobservable. 
We focus on the prevention efforts for self-protection because the interaction between market insurance and self-protection is inherently more complex than that with self-insurance, although our framework can be readily extended to examine market insurance and self-insurance interactions as well. 
Our analysis is within the risk minimization framework, where the insured's end-of-exposure risk is measured by the class of DRMs, which includes value-at-risk (VaR) and tail value-at-risk (TVaR) as prominent examples. 
DRMs are particularly suitable for this analysis due to their desirable mathematical properties including translation invariance, co-monotonic additivity, and positive homogeneity \citep{wang1997axiomatic}. 
Moreover, minimizing a DRM can be viewed as the dual of maximizing a utility function, with the dual utility characterized by a particular set of axioms outlined by \cite{yaari1987dual}. 
Rather focusing on the expected value premium principle, this paper adopts a broader class of convex premium principle to analyze the impact of the interaction between market insurance and self-protection. 
Additionally, we address the impact of moral hazard, which arises from asymmetric information in insurance models. 
In this context, moral hazard refers to the diminished incentive for insured individuals to take preventive actions, as insurance coverage reduces their need to minimize risks when insurers cannot directly observe their effort levels. 
We refer to \cite{winter2013optimal} and \cite{koohi2022systematic} for a comprehensive review of moral hazard in insurance.

The main contributions of the paper are summarized as follows: 

First, we revisit the optimal insurance demand and self-protection effort by minimizing a general DRM, 
where the insurance is priced using a broader class of convex premium principles. 
Unlike most previous studies, which primarily focus on a two-state loss model, our analysis considers a general loss model, providing a broader application to various risk scenarios. 
To solve the proposed model, we reformulate it as a two-stage optimization problem and derive explicit solutions to the inner problem. 
Our findings indicate that the optimal insurance demand can be full insurance, null insurance, or co-insurance, depending on the relationships among the insured’s risk measure, expected loss, and weighted expected loss under effort. 
Based on this insight, the outer minimization problem can be further reformulated as a piecewise function with three segments, each of which is convex, simplifying the analysis and solution process. 

Second, based on the general results for optimal insurance demand and the procedures for determining the optimal effort level, 
we explicitly characterize the optimal combination of prevention effort and insurance demand within a self-protection model 
when the insured employs TVaR and strictly convex DRMs, respectively. 
Moreover, we conduct extensive numerical experiments to illustrate these results. 
Our findings reveal that as the premium loading increases, a complementary relationship between market insurance and self-protection emerges, highlighting the dynamic interplay between these two risk management strategies.

Third, we extend the insurance demand and self-protection model to cases where the insured's precise level of protection effort is unobservable to the insurer, leading to ex ante moral hazard. 
To address this challenge, we establish sufficient conditions that link the optimal insurance demand with the optimal effort level based on the insured's incentive compatibility constraint. 
For a concrete case using the TVaR measure, we derive an explicit expression for this relationship. 
Furthermore, we provide a numerical example to illustrate the effects of moral hazard on both the optimal effort and insurance demand. 
Our results reveal that moral hazard significantly reduces the insured's prevention efforts while altering the optimal insurance structure.

The remainder of the paper is organized as follows: 
Section \ref{sec:prelim} introduces the problem setup. 
In Section \ref{sec:prob}, we 
present the general solution for the inner sub-problem and formulate the simplified optimal prevention effort problem studied in the sequel. 
Section \ref{sec:main} is dedicated to the exact solutions to the formulated model under TVaR and strictly convex DRMs, respectively. 
Section \ref{sec:numerical} provides a comparative static analysis of the interaction of market insurance share and optimal effort in self-protection. 
Section \ref{sec:moral} focuses on the impact of moral hazard on the interaction between self-protection and insurance demand. 
Finally, we conclude the paper in Section \ref{sec:conclusion}. 
All the proofs are delegated to the Appendices to facilitate reading.

\section{Problem setup}
\label{sec:prelim}

Let $(\Omega,\mathcal{F},\mathbb{P})$ be a given probability space, where $\Omega$ is a sample space with sigma-algebra $\mathcal{F}$ and $\mathbb{P}$ is a reference probability measure. 
A non-negative and essentially bounded random variable $X$ defined on $\Omega$, 
with $M=\esssup X$, denotes the aggregate loss confronted by the policyholder. 
Let $F_X(x)$ and $S_X(x)$ represent the cumulative distribution function (CDF) and survival function (SF) of $X$, respectively. 
That is, $F_X(x)=\mathbb{P}(X\leq x)$ and $S_X(x)=1-F_X(x)$. 
The inverse of $F_X$ is denoted by 
\[
F_X^{-1}(t): = \inf\{ x\in \R: F_X(x)\geq t\},\quad t\in (0,1]. 
\]
For convenience, we adopt the following notation throughout the paper. 
The indicator function is denoted by $\mathds{1}_{A}(s)$, where $\mathds{1}_{A}(s)=1$ for $s\in A$ and $\mathds{1}_{A}(s)=0$ for $s\notin A$. 
We use calligraphic letters for sets (e.g., $\mathcal{H}$). 


Consider a decision-maker (\textit{in short}, DM) faced with risk $X$ and purchasing insurance from an insurance company. 
In exchange for undertaking the ceded loss $I(X)$ from the insurer, the DM is compensated by the \emph{convex premium principle}: 
\begin{equation}
\label{expr:premium}
    \pi(I(X))=\E[h(I(X))],
\end{equation}
where $h:[0, M]\mapsto[0, M]$ is a non-decreasing and convex function satisfying $h(0)=0$ and $h(x)\geq x$ for any $x\in [0,M]$. 
This type of premium principle is recently used in \cite{cao2024optimal} and \cite{ghossoub2025bowley}. 
We adopt this broader class of premium principles due to their economic appeal. 
Specifically, these premium principles feature a marginal premium that increases with the size of the loss. 
We also refer interested readers to \cite{deprez1985convex} and \cite{kaluszka2005optimal} for other  convex premium principles in the literature. 
To simplify the analysis that follows, throughout the paper, we assume that the function $h$ in (\ref{expr:premium}) is strictly convex.



Next, we recall the well-known class of DRMs, which will be employed to quantify the end-of-period risk of the DM.

\begin{Def}\label{def:drm}
    A distortion function $g: [0, 1]\mapsto [0, 1]$ is a non-decreasing function with $g(0)=0$ and $g(1)=1$. 
    A \emph{DRM} $\rho_g$ of a nonnegative random variable $X$ with a distortion function $g$ is defined as
\begin{equation}
\rho_g(X)=\int_0^{\infty}g\left(S_X(t)\right) \dif t,
\end{equation}
provided that the integral is finite. 
\end{Def}

The set of all distortion functions is denoted by $\mathcal{G}$, and the subset of distortion functions that are concave on $[0, 1]$ is denoted by $\mathcal{G}_{cv}$.
It holds that $\mathcal{G}_{cv}\subset \mathcal{G}$. 
For any left-continuous distortion function $g$, define $\tilde{g}(s)=1-g(1-s)$ for $s\in [0,1]$, which is right-continuous. 
Then, the DRM $\rho_g$ can be equivalently expressed as 
\begin{equation}
\label{eq:alternative_DRM}
\rho_g(X)=\int_0^1 \mathrm{VaR}_{s}(X) d \tilde{g}(s) = \int_0^1 \mathrm{VaR}_{1-s}(x)d g(s)
\end{equation}
see, for example, \cite{dhaene2012remarks}. 
It is important to note that DRMs satisfy the following desirable properties, 
such as 
monotonicity, comonotonic additivity, translation invariance, positive homogeneity, and subadditivity (if $g\in\mathcal{G}_{cv}$).

Two prominent examples of DRMs are 
the VaR and the TVaR. 
The {\rm VaR} and {\rm TVaR} of a nonnegative random variable $X$ at a confidence level $\beta \in(0,1)$ are defined as $\mathrm{VaR}_{\beta}(X) =F_X^{-1}(\beta)$ 
and
\begin{equation*}
    \mathrm{TVaR}_{\beta}(X) = \frac{1}{1-\beta}\int_{\beta}^1 \mathrm{VaR}_{s}(X) \dif s,
\end{equation*}
provided that the integral exists. Indeed, {\rm VaR} and {\rm TVaR} are two DRMs, corresponding to the distortion functions $g(t)=\mathds{1}_{(1-\beta, 1]} (t)\in\mathcal{G}$ and $g(t)=\min\{1,t/(1-\beta)\}\in\mathcal{G}_{cv}$, respectively. 



%

For the self-protection model, we adopt the same setting as in \cite{bensalem2020prevention}, where 
the DM controls his loss probability by exerting an effort $e$ on it. 
More precisely, let the probability distribution $\mathbb{P}_{X_e}$ of the random variable $X_e$ take the following form
\begin{equation}
\label{eq:selfprotection}
\mathbb{P}_{X_e}=(1-p(e))\delta_{\{0\}}+p(e)\mathbb{P}_Y
\end{equation}
where $e\mapsto p(e)$ is a non-increasing function, $\delta_{\{0\}}$ is the Dirac mass at 0, and $\mathbb{P}_Y$ represents the distribution of a positive random variable $Y$, which takes values in the interval $[0,M]$. 
In other words, the loss random variable $X_e$ is strictly positive with probability $p(e)$, in which case its value is given by $Y$, and takes the value 0 with probability $1 - p(e)$. 
It is vital to note that the family of random variables $(X_e)_{e\in (0,+\infty)}$, indexed by a prevention effort $e$, decreases in the sense of the first-order stochastic dominance (FSD) as the effort $e$ increases. 
In other words, we have $X_{e_1}\succeq_{\mathrm{FSD}} X_{e_2}$ when $e_1\leq e_2$. 
Consequently, the mapping $e\mapsto \mathrm{VaR}_{\beta}(X_e)$ is non-increasing. 
In the following discussions, we assume that the protection probability $p(e)$ has the following properties:
\begin{ass}
\label{protection:e} 
Assume that $p(e)$ is non-increasing and strictly convex, and such that $$p_0:=p(0)\in(0,1),\quad p_{\infty}:=p(+\infty)=0,\quad p'(0)<0 ~\mbox{ and }~p'(+\infty)=0.$$
\end{ass}
%
Assumption \ref{protection:e} is standard in insurance economics, see, e.g., \cite{courbage2013prevention}, which means that an increased effort leads to a decrease in the loss probability $p(e)$, with a decreasing marginal impact of the effort. 
$p_0\in (0,1)$ means the existence of the original risk of the DM with no effort, while $p_{\infty}=0$ represents that the original risk can be fully eliminated.\footnote{It is remarkable to mention that all the results presented in this paper can be extended to the case of $p_{\infty}\in(0,p_0)$.} 
Moreover, the condition $\quad p'(0)<0$ means that the initial effort of prevention is effective and condition $p'(+\infty)=0$ means that the marginal effect becomes insignificant
when the effort $e$ approaches infinite. 



Consider a DM facing a loss risk $X_e$ characterized by the distribution specified in \eqref{eq:selfprotection}. 
To mitigate this risk, the DM purchases a proportional insurance contract from an insurer, selecting the coverage level $\alpha\in [0,1]$. 
The associated insurance premium is denoted by $\pi(\alpha X_e)$.
In addition to purchasing insurance, the policyholder can exert a prevention effort $e\in \R_+$, incurs a monetary cost $c(e)$. 
%
%
The policyholder’s objective is to minimize the risk measure associated with their total loss, which is given by: 
\begin{equation}
\label{prob:original}
\min_{(e,\alpha)\in\mathbb{R}_{+}\times[0,1]}\rho_{g}\left(X_e-\alpha X_e+\mathbb{E}[h(\alpha X_e)]+c(e)\right).
\end{equation}
Note that the proposed model (\ref{prob:original}) is one-period model in the principal-agent literature. We also refer interested readers to \cite{liu2023optimal} for a continuous-time model that studies a similar problem. 
Using the translation invariance and positive homogeneity of DRM, the objective function in (\ref{prob:original}) can be simplified to 
$\mathbb{E}[h(\alpha X_e)] - \alpha\rho_{g}\left(X_e\right) + \rho_g(X_e) +c(e)$, 
which is convex in $\alpha$ for any fixed $e\in \mathbb{R}_+$ since $h$ is a convex function. 
Consequently, the minimization problem 
\[
\min_{\alpha\in[0,1]}\rho_{g}\left(X_e-\alpha X_e+\mathbb{E}[h(\alpha X_e)]+c(e)\right)
\]
is well-behaved and achieves the global minimum over $\alpha\in [0,1]$ for any fixed $e\in \mathbb{R}_+$. 
Therefore, problem (\ref{prob:original}) 
can be solved sequentially by optimizing
\begin{equation}\label{prob:innerouter}
\min_{e\in\mathbb{R}_{+}}\left\{\min_{\alpha\in[0,1]}\rho_{g}\left(X_e-\alpha X_e+\mathbb{E}[h(\alpha X_e)]+c(e)\right)\right\}.
\end{equation}

To facilitate subsequent discussions, 
we assume that the cost function $c(e)$ satisfies the following assumption according to the marginal impact of the effort:
\begin{ass}
\label{cost:e} 
Assume that the cost function $c(e)$ is continuously differentiable, non-decreasing, and strictly convex, and satisfies the following conditions:
\[
c(0)=0,\quad c(+\infty)=+\infty,\quad c'(0)=0, ~\mbox{ and }~c'(+\infty)=+\infty.
\]
\end{ass}

This assumption asserts that no prevention effort ($e=0$) incurs no monetary cost, and even minimal prevention efforts can significantly reduce risk. 
However, achieving complete prevention requires infinite effort, which corresponds to an infinite cost.
Furthermore, the strict convexity of $c(e)$ ensures that the marginal cost of effort increases with higher levels of investment. This reflects the realistic principle that reducing risk becomes progressively more expensive as prevention efforts intensify. 




\section{General solutions}
\label{sec:prob}
We now deal with the inner optimization problem of problem \eqref{prob:innerouter}. 
By the translation invariance and positive homogeneity of DRM, we have 
\begin{eqnarray}
\label{prob:funcalpha}
  \rho_{g}\left(X_e-\alpha X_e+\mathbb{E}[h(\alpha X_e)]+c(e)\right) 
  =
  \mathbb{E}[h(\alpha X_e)]-\alpha\rho_{g}\left(X_e\right)+\rho_{g}\left(X_e\right)+c(e).
\end{eqnarray}
Let $J(\alpha)=\mathbb{E}[h(\alpha X_e)]-\alpha\rho_{g}\left(X_e\right)$. 
It is clear that\footnote{The function $h$ is strictly convex, so $h$ is twice differentiable almost everywhere, see, e.g., \cite{fu2011extension}. 
}
\begin{equation*}
    J'(\alpha)=\mathbb{E}[X_eh'(\alpha X_e)]-\rho_{g}\left(X_e\right)
\end{equation*}
and
\begin{equation*}
    J''(\alpha)=\mathbb{E}[X_e^2h''(\alpha X_e)]\geq0. 
\end{equation*}
Thus, $J(\alpha)$ is convex in $\alpha\in[0,1]$. Note that $J'(\alpha)$ is non-decreasing in $\alpha\in[0,1]$. 
Therefore, to minimize $J(\alpha)$ over $\alpha\in[0,1]$, we consider the following three cases:
\begin{itemize}
    \item If $J'(1)\leq0$, i.e. $\mathbb{E}[X_eh'(X_e)]\leq\rho_{g}\left(X_e\right)$, then $J'(\alpha)\leq J'(1)\leq0$ for all $\alpha\in[0,1]$. Hence, the minimum value of $J(\alpha)$ is achieved at $\alpha=1$. 
    \item If $J'(0)\geq0$, i.e. $h'(0)\mathbb{E}[X_e]\geq\rho_{g}\left(X_e\right)$, then $J'(\alpha)\geq J'(0)\geq0$ for all $\alpha\in[0,1]$. 
    Hence, the minimum value of $J(\alpha)$ is achieved at $\alpha=0$.
    \item If $J'(1)>0>J'(0)$, i.e. $h'(0)\mathbb{E}[X_e]<\rho_{g}\left(X_e\right)<\mathbb{E}[X_eh'(X_e)]$, then $J'(\alpha)=0$ must give an unique solution on $\alpha\in(0,1)$. 
    That is, $\alpha_{h,e}^*\in(0,1)$ is determined from
    \begin{equation}
    \label{eq:alpha_h}
        \mathbb{E}[X_eh'(\alpha X_e)]=\rho_{g}\left(X_e\right),\quad \alpha\in(0,1).
    \end{equation}
\end{itemize}

To sum up, the solution (denoted by $\alpha^*_e$) of the inner minimization problem of problem \eqref{prob:funcalpha} can be denoted as follows:
\begin{equation}\label{solution:alpha}
    \alpha^*_e=\mathds{1}_{\{\mathbb{E}[X_eh'(X_e)]\leq\rho_{g}\left(X_e\right)\}}+\alpha_{h,e}^*\mathds{1}_{\{h'(0)\mathbb{E}[X_e]<\rho_{g}\left(X_e\right)<\mathbb{E}[X_eh'(X_e)]\}}.
\end{equation}
Substituting \eqref{solution:alpha} into \eqref{prob:innerouter}, the original problem boils down to solving
\begin{equation}\label{prob:outer}
\min_{e\in\mathbb{R}_{+}}\mathbb{E}[h(\alpha^*_e X_e)]-\alpha^*_e\rho_{g}\left(X_e\right)+\rho_{g}\left(X_e\right)+c(e).
\end{equation}

Note that the value of $\alpha^*_e$ depends on the relations among the three terms $h'(0)\mathbb{E}[X_e]$, $\rho_{g}\left(X_e\right)$ and $\mathbb{E}[X_eh'(X_e)]$.\footnote{It should be noted that $\mathbb{E}[X_eh'(X_e)]$ cannot be treated as a DRM. In the work of \cite{bensalem2020prevention}, the authors considered the expected value premium principle, i.e. $h'(\cdot)=1+\theta$. 
In our problem, we exclude the case of the expected value premium principle by assuming that $h'(\cdot)$ is strictly increasing. Hence, as seen later, our results are very different with \cite{bensalem2020prevention}.}  
Since $X_{e_1}\succeq_{\rm FSD}X_{e_2}$ for any $e_1\leq e_2$, 
it follows that all of the three terms are non-increasing functions in $e\in\mathbb{R}_+$. 
Since $h'(0)\mathbb{E}[X_e]\leq\mathbb{E}[X_eh'(X_e)]$, we only need to discuss the magnitude of  $\rho_{g}\left(X_e\right)$ and $h'(0)\mathbb{E}[X_e]$, and $\rho_{g}\left(X_e\right)$ and $\mathbb{E}[X_eh'(X_e)]$. Define
\begin{equation}\label{def:A1A2}
    \mathcal{A}_1=\left\{e\in\mathbb{R}_+\mid h'(0)\geq \frac{\rho_{g}\left(X_e\right)}{\mathbb{E}[X_e]}\right\}\quad\mbox{and}\quad\mathcal{A}_2=\left\{e\in\mathbb{R}_+\mid \frac{\rho_{g}\left(X_e\right)}{\mathbb{E}[X_eh'(X_e)]}\geq1\right\}.
\end{equation}
Denote by
\begin{equation}\label{def:G}
        G_1(e):=\frac{\rho_{g}\left(X_e\right)}{\mathbb{E}[X_e]}\quad\mbox{and}\quad G_2(e):=\frac{\rho_{g}\left(X_e\right)}{\mathbb{E}[X_eh'(X_e)]}.
\end{equation}
The two sets $\mathcal{A}_1$ and $\mathcal{A}_2$ can be simplified into
\begin{equation}\label{expr:A1A2}
    \mathcal{A}_1=\left\{e\in\mathbb{R}_+\mid h'(0)\geq G_1(e)\right\}\quad\mbox{and}\quad\mathcal{A}_2=\left\{e\in\mathbb{R}_+\mid G_2(e)\geq1\right\}.
\end{equation}
Note that $\mathcal{A}_2\subseteq \mathcal{A}_1^c$, it follows that $\alpha^*_e=\mathds{1}_{\mathcal{A}_2}+\alpha_{h,e}^*\mathds{1}_{\mathcal{A}_1^c\cap\mathcal{A}_2^c}$. 
Denote by $$K(e):=\mathbb{E}[h(\alpha^*_e X_e)]-\alpha^*_e\rho_{g}\left(X_e\right)+\rho_{g}\left(X_e\right)+c(e)$$ for the objective function in \eqref{prob:outer}. 
It is essential to note that the function $e\mapsto K(e)$ is not convex, so its minimization on $\R_+$ is not straightforward. 
However, we can take advantage of the fact that $\alpha_e^*$ takes the values $0, 1$ and $\alpha_{h,e}^*$ to study $K(e)$ on the sets $\mathcal{A}_1$, $\mathcal{A}_2$ and $\mathcal{A}_1^c\cap\mathcal{A}_2^c$, separately, 
and then compare the local minima on theses sets to get a global minimum. 
Specifically, we have 
\begin{equation}\label{expr:Ke}
    K(e)=\begin{cases}
        \rho_{g}\left(X_e\right)+c(e), &\quad \mathrm{if}\ e\in \mathcal{A}_1;\\
        \mathbb{E}[h(\alpha_{h,e}^* X_e)]+(1-\alpha_{h,e}^*)\rho_{g}\left(X_e\right)+c(e), &\quad \mathrm{if}\ e\in\mathcal{A}_1^c\cap\mathcal{A}_2^c;\\
        \mathbb{E}[h(X_e)]+c(e), &\quad \mathrm{if}\ e\in \mathcal{A}_2.
    \end{cases}
\end{equation}
Henceforth, we will consider minimizing $K(e)$ given in \eqref{expr:Ke} and finding the optimal protection effort $e^*$ under the framework of convex DRMs.\footnote{For the case of VaR measure, we can also obtain the optimal solutions on insurance demand and protection effort; however, the discussions are very complex, and the results can be shown to the readers upon request.} 
%


\section{Main results}
\label{sec:main}

In this section, we explicitly characterize the optimal combination of prevention effort and insurance demand in a self-protection model when the insured adopts TVaR measure and strictly convex distortion risk measures, respectively. 

\subsection{TVaR measure}
\label{sec:4.1}

We investigate the optimal proportional insurance demand and protection effort under the TVaR measure with $g(t)=\min\{1,t/(1-\beta)\}$. 
According to \eqref{eq:selfprotection}, the distribution of $X_e$ has the Dirac mass at 0 with probability $1-p(e)$ and has a continuous positive loss $Y$ with probability $p(e)$. It is easy to calculate the expression of the VaR measure of $X_e$ as follows: 
\begin{equation}
\label{varexpression}
    \text{VaR}_\beta(X_e)=\begin{cases}
        0, & 0<\beta\leq 1-p(e);\\
        \text{VaR}_{\frac{\beta+p(e)-1}{p(e)}}(Y), & 1-p(e)<\beta<1,
    \end{cases}
\end{equation}
Define $$\beta(e):=\frac{\beta+p(e)-1}{p(e)}\quad\text{ and }\quad e_\beta:=p^{-1}(1-\beta).$$
According to Assumption \ref{protection:e}, we know that $p(e)$ is a non-increasing and strictly convex function of $e$, it thus follows that $\beta(e)$ is a non-increasing continuous function of $e$ and the inverse of $p$ (i.e., $p^{-1}(\cdot)$) is well-defined. 
Note that
$$
    \text{TVaR}_{\beta}(X_e)=\frac{1}{1-\beta}\int_{\beta}^1\text{VaR}_s(X_e)\dif s=\begin{cases}
        \frac{1}{1-\beta}\int_{\beta}^{1}\text{VaR}_s(X_e)\dif s, &0\leq e<e_{\beta};\\
        \frac{1}{1-\beta}\int_{1-p(e)}^{1}\text{VaR}_s(X_e)\dif s, &e\geq e_{\beta}.
    \end{cases}
$$
Based on \eqref{varexpression}, the above expression of TVaR can be simplified as follows:
\begin{align}\label{TVaR:express}
    \text{TVaR}_\beta(X_e)=&\begin{cases}
    \frac{1}{1-\beta}\int_\beta^1\text{VaR}_{\frac{s+p(e)-1}{p(e)}}(Y) \dif s, & 0\leq e<e_\beta;\\
    \frac{1}{1-\beta}\int_{1-p(e)}^1 \text{VaR}_{\frac{s+p(e)-1}{p(e)}}(Y) \dif s,& e\geq e_\beta,
    \end{cases}\nonumber\\
    =&\begin{cases}
        \frac{p(e)}{1-\beta}\int_{\beta(e)}^1\text{VaR}_t(Y) \dif t, &0\leq e<e_\beta;\\
        \frac{p(e)}{1-\beta}\mathbb{E}[Y], &e\geq e_\beta.
    \end{cases}
\end{align}
Since $p(e)$ and $\beta(e)$ are both continuous in $e\in\mathbb{R}_+$, and $\beta(e_\beta)=0$, we have $\text{TVaR}_\beta(X_e)$ is also continuous in $e\in\mathbb{R}_+$ 
by noting that $\mathbb{E}[Y]=\int_{0}^1\text{VaR}_t(Y) \dif t$.

In light of the above analysis, the expressions of $G_1$ and $G_2$, which are defined in \eqref{def:G}, can be simplified as
\begin{equation}\label{expr:G1}
    G_1(e)=\begin{cases}
        \frac{1}{(1-\beta)\mathbb{E}[Y]}\int_{\beta(e)}^1\text{VaR}_s(Y) \dif s, &0\leq e<e_\beta;\\
        \frac{1}{1-\beta}, &e\geq e_\beta.
    \end{cases}
\end{equation}
and
\begin{equation}\label{expr:G2}
    G_2(e)=\begin{cases}
        \frac{1}{(1-\beta)\mathbb{E}[Yh'(Y)]}\int_{\beta(e)}^1\text{VaR}_s(Y) \dif s, &0\leq e<e_\beta;\\
        \frac{\mathbb{E}[Y]}{(1-\beta)\mathbb{E}[Y h'(Y)]}, &e\geq e_\beta.
    \end{cases}
\end{equation}
It is obvious to note that both $G_1(e)$ and $G_2(e)$ are non-decreasing in $e\in[0,e_\beta)$ and remain as constants over $[e_\beta,+\infty)$.

By applying the strict convexity of $h$ (or equivalently, the strict increasingness of $h'$), it follows that $G_1(e)> G_2(e)$, for all $e\in\mathbb{R}_+$. Let $\beta_0:=\beta(0)=\frac{\beta+p_0-1}{p_0}$. It is clear that $\beta(e)\in(0,\beta_0]$ for all $e\in[0,e_\beta)$. Furthermore, note that  $$G_1(0)=\frac{\int_{\beta_0}^1\text{VaR}_s(Y)\dif s}{(1-\beta)\mathbb{E}[Y]} ~\text{ and } ~G_2(0)=\frac{\int_{\beta_0}^1\text{VaR}_s(Y)\dif s}{(1-\beta)\mathbb{E}[Y h'(Y)]}.$$
Besides, under the TVaR measure, we can give a more accurate expression of equation \eqref{eq:alpha_h} for which $\alpha_{h,e}^*$ is satisfied:
\begin{equation}\label{expr:alphahe}
\int_{\beta(e)}^1\text{VaR}_s(Y)\dif s=(1-\beta)\mathbb{E}[Y h'(\alpha_{h,e}^*Y)],~ ~\text{ for }~ ~e\in[0,e_\beta].
\end{equation}

The following theorem summarizes the explicit solutions of $\alpha_e^*$ under the TVaR measure.
\begin{thm}
\label{thm:alphae}
Under the TVaR measure and Assumption \ref{protection:e}, the solution of inner optimization problem of \eqref{prob:innerouter} is given as follows:
    \begin{enumerate}[(i)]
        \item If $G_1(0)>h'(0)$ and $G_2(0)>1$, then $\alpha_e^*\equiv1$ for all $e\in\mathbb{R}_+$.
        \item If $G_1(0)>h'(0)$ and $G_2(0)\leq 1\leq G_2(e_\beta)$, then 
        $$\alpha_e^*=\begin{cases}
        \alpha_{h,e}^*, &e\in[0,e_{G_2});\\
        1, &e\in[e_{G_2},+\infty),
        \end{cases}$$ 
        where $e_{G_2}$ is the solution of $G_2(e)=1$ on $[0,e_\beta]$.
        \item If $G_1(0)>h'(0)$ and $G_2(e_\beta)<1$, then 
        $$\alpha_e^*=\begin{cases}
            \alpha_{h,e}^*, &e\in[0,e_\beta);\\
            \alpha_{h,e_\beta}^*, &e\in[e_\beta,+\infty),
        \end{cases}$$
        where $\alpha_{h,e_\beta}^*$ is the solution of $\mathbb{E}[Y]=(1-\beta)\mathbb{E}[Yh'(\alpha Y)]$ for $\alpha\in(0,1)$.
        \item If $G_1(0)\leq h'(0)\leq G_1(e_\beta)$, $G_2(0)\leq1\leq G_2(e_\beta)$, then $$\alpha_e^*=\begin{cases}
        0, &e\in[0,e_{G_1}];\\
        \alpha_{h,e}^*, &e\in(e_{G_1},e_{G_2});\\
        1, &e\in[e_{G_2},+\infty),
        \end{cases}$$
        where $e_{G_1}$ is the solution of $G_1(e)=h'(0)$ on $[0,e_\beta]$.
        \item If $G_1(0)\leq h'(0)\leq G_1(e_\beta)$ and $G_2(e_\beta)<1$, then $$\alpha_e^*=\begin{cases}
            0, &e\in[0,e_{G_1}];\\
            \alpha_{h,e}^*, &e\in(e_{G_1},e_\beta);\\
            \alpha_{h,e_\beta}^*, &e\in[e_\beta,+\infty).
        \end{cases}$$
        \item If $G_1(e_\beta)<h'(0)$ and  $G_2(e_\beta)<1$, then $\alpha_e^*\equiv0$ for all $e\in\mathbb{R}_+$.
    \end{enumerate}
Moreover, for cases (ii)-(v), the solution $\alpha_{h,e}^*$ is continuous and non-decreasing in $e$ with respect to the corresponding intervals. Therefore, the solution $\alpha_e^*$ is continuous and non-decreasing in $e\in\mathbb{R}_+$, for all cases.
\end{thm}

As demonstrated in Theorem \ref{thm:alphae}, there are six distinct cases for the solution of $\alpha_e^*$ corresponding to a given $e\in \R_+$. 
We substitute the results from each case into equation (\ref{expr:Ke}) and optimize the objective function $K(e)$ in each scenario to determine the optimal protection effort $e^*$. 
For simplicity, we denote the objective function $K(e)$ under each case in Theorem \ref{thm:alphae} by $K_j(e)$ for $j=1,\ldots,6$. 
%
Before presenting the results for the optimal protection effort $e^*$, we introduce the following two assumptions to guarantee the uniqueness of the local minimum points for the different intervals. 

\begin{ass}
\label{ass:convexTVaR} 
For any $\beta\in (0,1)$, the function $e\mapsto \mathrm{VaR}_{\beta}(X_e)$ is convex on the interval $[0,e_{\beta}]$. 
\end{ass}
Assumption \ref{ass:convexTVaR} is adopted from \cite{bensalem2020prevention}, which means that the marginal impact of effort on the VaR position is non-decreasing. 
Under Assumption \ref{ass:convexTVaR}, it is easy to verify that the function $e\mapsto \rho_g(X_e)$ in \eqref{eq:alternative_DRM} is non-decreasing and convex provided that $g$ is concave, which will be utilized in Section \ref{sec:moral} for studying the problem under ex ante moral hazard. 
In the following, we provide a sufficient condition to ensure that the assumption \ref{ass:convexTVaR} holds.

\begin{rmk}
Assume that $Y$ is a continuous positive random variable with a probability distribution function $f_Y(y)$ defined for $y\in [0,M]$. 
    Let $y:=F_Y^{-1}\left(\frac{\beta+p(e)-1}{p(e)}\right)=\mathrm{VaR}_{\frac{\beta+p(e)-1}{p(e)}}(Y)$ for $e\in[0,e_{\beta}]$. 
    Then we have $\mathrm{VaR}_{\beta}(X_e)=y$ for $e\in[0,e_{\beta}]$ (see (\ref{varexpression})). 
    Taking the first derivative of $y$ w.r.t. $e$ gives rise to
    \[
    \frac{\dif y}{\dif e}=\frac{(1-\beta)p'(e)}{p^2(e)\cdot f_Y(y)}.
    \]
    The second derivative of $y$ w.r.t. $e$ can be computed as
    \[
    \frac{\dif^2y}{\dif e^2}=\frac{(1-\beta)p(e)f_Y^2(y)[p''(e)p(e)-2(p'(e))^2]-(1-\beta)^2(p'(e))^2f'_Y(y)}{p^4(e)\cdot f_Y^3(y)}.
    \]
    If $p''(e)p(e)\geq2(p'(e))^2$ and $Y$ has a decreasing density function, 
    then Assumption \ref{ass:convexTVaR} holds. It is worth mentioning that the condition $p''(e)p(e)\geq2(p'(e))^2$ is commonly used in many insurance economics literature such as \cite{jullien1999should} and \cite{snow2011ambiguity}. 
    Besides, there are many distribution functions having decreasing probability density functions including the exponential distribution, the Gamma and Weibull distributions with shape parameters smaller than 1, and the Pareto distribution.
\end{rmk}


\begin{ass}
\label{ass:uniqueTVaR} 
The minimum point of $K(e)$ is unique on $e\in\mathcal{A}_1^c\cap\mathcal{A}_2^c$. 
\end{ass}
Assumption \ref{ass:uniqueTVaR} is a standard technical condition commonly used to ensure uniqueness and well-posedness in optimization problems. 
The following theorem summarizes the optimal prevention effort $e^*$ based on the results from Theorem \ref{thm:alphae}.

\begin{thm}
\label{thm:e}
    Under the settings of Theorem \ref{thm:alphae} and Assumptions \ref{cost:e}, 
    \ref{ass:convexTVaR} 
    and \ref{ass:uniqueTVaR}, the optimal effort $e^*$ is summarized as follows:
    \begin{enumerate}[(i)]
        \item If $G_1(0)>h'(0)$ and $G_2(0)>1$, then $e^*=\hat{e}$, where $\hat{e}$ is the solution to $K_1'(e)=0$ for $e\in\mathbb{R}_+$ with $K_1(e)$ given in (\ref{expr:K1}).
        \item If $G_1(0)>h'(0)$ and $G_2(0)\leq 1\leq G_2(e_\beta)$, then $K_2(e)$ has the form (\ref{expr:K2}), and the optimal $e^*$ can be solved as follows: if there exists an interior point $\tilde{e}$ such that  $\tilde{e}=\arg\min_{e\in[0,e_{G_2})}K_2(e)$, $K_2(\tilde{e})\leq K_2(\hat{e})$, then $e^*=\tilde{e}$; otherwise, $e^*=\hat{e}$, where $\hat{e}$ is the minimum point of $K_2(e)$ for $e\in[e_{G_2},+\infty)$.
        \item If $G_1(0)>h'(0)$ and $G_2(e_\beta)<1$, then $K_3(e)$ has the form (\ref{expr:K3}), and the optimal $e^*$ can be derived as follows: if there exists an interior point $\tilde{e}$ such that  $\tilde{e}={\arg\min}_{e\in[0,e_\beta)}K_3(e)$ and $K_3(\tilde{e})\leq K_3(\text{\r e})$, then $e^*=\tilde{e}$; otherwise, $e^*=\text{\r e}$, where $\text{\r e}$ is the minimum point of $K_3(e)$ for $e\in[e_\beta,+\infty)$.    
        \item If $G_1(0)\leq h'(0)\leq G_1(e_\beta)$ and $G_2(0)\leq1\leq G_2(e_\beta)$, then $K_4(e)$ has the form (\ref{expr:K4}), and 
        \begin{enumerate}[(a)]
            \item if there exists an interior point $\tilde{e}$ such that  $\tilde{e}={\arg\min}_{e\in(e_{G_1},e_{G_2})}K_4(e)$ and $K_4(\tilde{e})\leq \min\{K_4(\Dot{e}),K_4(\hat{e})\}$, then $e^*=\tilde{e}$;
            \item otherwise, if $K_4(\Dot{e})\leq K_4(\hat{e})$, then $e^*=\Dot{e}$;
            \item otherwise, if $K_4(\Dot{e})> K_4(\hat{e})$, then $e^*=\hat{e}$,
        \end{enumerate}
        where $\Dot{e}$ is the minimum point of $K_4(e)$ for $e\in[0,e_{G_1}]$ and $\hat{e}$ is the minimum point of $K_4(e)$ for $e\in[e_{G_2},+\infty)$.
        \item If $G_1(0)\leq h'(0)\leq G_1(e_\beta)$ and $G_2(e_\beta)<1$, then $K_5(e)$ has the form (\ref{expr:K5}), and
        \begin{enumerate}[(a)]
            \item if there exists an interior point $\tilde{e}$ such that  $\tilde{e}={\arg\min}_{e\in(e_{G_1},e_\beta)}K_5(e)$ and $K_5(\tilde{e})\leq \min\{K_5(\Dot{e}),K_5(\text{\r e})\}$, then $e^*=\tilde{e}$;
            \item otherwise, if $K_5(\Dot{e})\leq K_5(\text{\r e})$, then $e^*=\Dot{e}$;
            \item otherwise, if $K_5(\Dot{e})> K_5(\text{\r e})$, then $e^*=\text{\r e}$,
        \end{enumerate}
        where $\Dot{e}$ is the minimum point of $K_5(e)$ for $e\in[0,e_{G_1}]$ and $\text{\r e}$ is the minimum point of $K_5(e)$ for $e\in[e_\beta,+\infty)$.
        \item If $G_1(e_\beta)<h'(0)$ and $G_2(e_\beta)<1$, then $K_6(e)$ has the form (\ref{expr:K6}), and  if $K_6(\Dot{e})\leq K_6(\hat{e})$, then $e^*=\Dot{e}$; otherwise $e^*=\hat{e}$, where $\Dot{e}$ is the minimum point on $[0,e_\beta]$ and $\hat{e}$ is the minimum point on $[e_\beta,+\infty)$.
    \end{enumerate}
\end{thm}

By substituting the solutions from Theorem \ref{thm:e} into Theorem \ref{thm:alphae}, the optimal insurance policy and prevention strategy, $(\alpha^*_{e^*},e^*)$, can be immediately determined. 
The detailed calculations and economic interpretations will be presented in Section \ref{sec:numerical} as part of the numerical studies.

\subsection{Strictly convex distortion risk measure}\label{sec:convexDRM}

This section considers the optimal protection effort for the class of DRMs with strictly concave distortion functions. 
Note that, for a nonnegative random variable $Z$ and distortion function $g$, one has
\begin{equation*}
\rho_g(Z)=\int_0^{\infty}g\left(S_Z(t)\right) \dif t=\int_0^1 \text{VaR}_{1-u}(Z)\dif g(u).
\end{equation*}
According to the self-protection distribution model in \eqref{eq:selfprotection}, it is easy to calculate that
$$\text{VaR}_{1-u}(X_e)=\begin{cases}
    \text{VaR}_{1-\frac{u}{p(e)}}(Y)=F_Y^{-1}\left(1-\frac{u}{p(e)}\right), & 0< u<p(e),\\
    0, &p(e)\leq u<1.
\end{cases}$$
Then, $\rho_g(X_e)$ has the following expression 
\begin{eqnarray}
\label{convex:generalexpre}
\rho_g(X_e)&=&\int_0^1 \text{VaR}_{1-u}(X_e) \dif g(u)=\int_0^{p(e)} \mbox{VaR}_{1-\frac{u}{p(e)}}(Y) \dif g(u)\nonumber\\
&=&\int_0^{p(e)} F_Y^{-1}\left(1-\frac{u}{p(e)}\right) \dif g(u). 
\end{eqnarray} 
For simplicity, we denote by 
\[
\psi(e):=\frac{\int_0^{p(e)} \text{VaR}_{1-u}(X_e) \dif g(u)}{p(e)}=\frac{\int_0^{p(e)} F_Y^{-1}\left(1-\frac{u}{p(e)}\right) \dif g(u)}{p(e)}.
\]
Consequently, the expressions of $G_1$ and $G_2$, which are defined in (\ref{def:G}), can be simplified as
\begin{equation}
\label{eq:G1_strict}
    G_1(e)=\frac{\int_0^{p(e)} \mbox{VaR}_{1-\frac{u}{p(e)}}(Y) \dif g(u)}{p(e)\mathbb{E}[Y]}=\frac{\psi(e)}{\mathbb{E}[Y]}
\end{equation}
and
\begin{equation}
    G_2(e)=\frac{\int_0^{p(e)} \mbox{VaR}_{1-\frac{u}{p(e)}}(Y) \dif g(u)}{p(e)\mathbb{E}[Yh'(Y)]}=\frac{\psi(e)}{\mathbb{E}[Yh'(Y)]}. 
\end{equation}
By applying the strict convexity of $h$, it follows that $G_1(e)>G_2(e)$ for all $e\in\mathbb{R}_+$. 
Note that $$G_1(0)=\frac{\psi(0)}{\mathbb{E}[Y]}~\text{and}~ G_2(0)=\frac{\psi(0)}{\mathbb{E}[Yh'(Y)]}.$$
Furthermore, under strictly convex DRM, equation \eqref{eq:alpha_h} can be reformulated as 
\begin{equation}
\label{expr:alphahe2}
    p(e)\mathbb{E}[Y\cdot h'(\alpha_{h,e}^*Y)]=\int_0^{p(e)}\text{VaR}_{1-\frac{u}{p(e)}}(Y)\dif g(u), \text{ for all } e\in\mathbb{R}_+,
\end{equation}
provided that the above equation is well-defined.

The following assumption, resorting to the well-known \textit{star-shaped order} denoted by $\preceq_\star$ to characterize the variability \citep[][]{jones2003empirical,belzunce2012comparison} of the underlying risk shaped by effort,\footnote{The star-shaped order is one of the partial orders which are scale invariant. It is known that $X \preceq_{\star} Y$ implies $\gamma_X \leq \gamma_Y$, where $\gamma_X=\sqrt{\operatorname{Var}[X]} / \mathbb{E}[X]$ and $\gamma_Y=\sqrt{\operatorname{Var}[Y]} / \mathbb{E}[Y]$ denote the coefficients of variation of $X$ and $Y$. Interested readers can refer to \cite{shaked2007stochastic} for more detailed treatment.} is necessary to facilitate our subsequent discussions.
\begin{ass}\label{ass:star}
\label{ass:psi} $X_{e_1}\preceq_\star X_{e_2}$ when $e_1\leq e_2$. 
\end{ass}

With Assumptions \ref{protection:e} and \ref{ass:psi} imposed on the self-protection model, it can be seen that the prevention effort not only alters the loss in the first-order stochastic dominance, but also shapes the variability of the loss. This changing pattern assumption has not  been discovered in the existing literature.
Observe that
\begin{equation*}
    G_1(e_1)=\frac{\rho_{g}\left(X_{e_1}\right)}{\mathbb{E}[X_{e_1}]}=\frac{\int_0^1 F_{X_{e_1}}^{-1}(t) \dif \tilde{g}(t)}{\int_0^1 F_{X_{e_1}}^{-1}(t) \dif t}.
\end{equation*}
Since $\tilde{g}(t)$ is convex, then from Theorem 3.25 of \cite{belzunce2012comparison}, it follows that $G_1(e_1)\leq G_1(e_2)$ under Assumption \ref{ass:psi}. 
Hence, $G_1(e)$ is a non-decreasing function. 
Furthermore, under the self-protection model \eqref{eq:selfprotection}, it follows that
\begin{equation*}
G_2(e)=\frac{\rho_{g}\left(X_e\right)}{\mathbb{E}[X_eh'(X_e)]}=G_1(e)\times\frac{\mathbb{E}[X_{e}]}{\mathbb{E}[X_eh'(X_e)]}=\frac{\mathbb{E}[Y]}{\mathbb{E}[Yh'(Y)]}G_1(e),
\end{equation*}
which means that $G_2(e)$ is also 
non-decreasing in $e\in\mathbb{R}_+$. 


The following theorem provides the explicit solution of $\alpha_e^*$ for the inner minimization of problem \eqref{prob:funcalpha} when the DRM is the strictly convex DRM.

\begin{thm}\label{thm:alphae2}
    Suppose that $g$ is strictly concave and Assumptions \ref{protection:e} and \ref{ass:psi} hold. 
    The optimal solution of the inner optimization problem of problem \eqref{prob:innerouter} is given as follows:
    \begin{enumerate}[(i)]
        \item If $G_1(0)>h'(0)$ and $G_2(0)>1$, then $\alpha_e^*\equiv1$ for all $e\in\mathbb{R}_+$.
        \item If $G_1(0)>h'(0)$ and $G_2(0)\leq1$, then $$\alpha_e^*=\begin{cases}
        \alpha_{h,e}^*, &e\in[0,e_{G_2});\\
        1, &e\in[e_{G_2},+\infty),
        \end{cases}$$ 
        where $e_{G_2}$ is the solution of $G_2(e)=1$ on $e\in\mathbb{R}_+$.
        \item If $G_1(0)\leq h'(0)$ and $G_2(0)\leq1$, then $$\alpha_e^*=\begin{cases}
        0, &e\in[0,e_{G_1}];\\
        \alpha_{h,e}^*, &e\in(e_{G_1},e_{G_2});\\
        1, &e\in[e_{G_2},+\infty),
        \end{cases}$$
        where $e_{G_1}$ is the solution of $G_1(e)=h'(0)$ on $e\in\mathbb{R}_+$.
    \end{enumerate}
Moreover, in cases (ii) and (iii), the solution $\alpha_{h,e}^*$ is continuous and non-decreasing in $e$ within the corresponding intervals. 
Finally, the solution $\alpha_e^*$ is continuous and non-decreasing in $e\in\mathbb{R}_+$ for all cases.
\end{thm}




As shown in Theorem \ref{thm:alphae2}, there are in total three cases on the solution of $\alpha_e^*$ for given $e\in\mathbb{R}_+$. We substitute the results into (\ref{expr:Ke}) and optimize the objective function $K(e)$ in each case to obtain the optimal protection effort $e^*$. Denote by $K_j(e)$ as the objective function $K(e)$ given in \eqref{expr:Ke} under each case in Theorem \ref{thm:alphae2}, for $j=1,2,3$. 
The following theorem summarizes the optimal prevention effort $e^*$ under the results of Theorem \ref{thm:alphae2}.
\begin{thm}\label{thm:e2}
    Under the settings of Theorem \ref{thm:alphae2} and Assumptions \ref{cost:e}, \ref{ass:uniqueTVaR} and \ref{ass:convexTVaR}, the optimal effort is summarized as follows:
    \begin{enumerate}[(i)]
        \item If $G_1(0)>h'(0)$ and $G_2(0)>1$, then $e^*=\hat{e}$, where $\hat{e}$ is the solution to $K_1'(e)=0$ for $e\in\mathbb{R}_+$ with $K_1(e)$ given in \eqref{expr:K1}.
        \item If $G_1(0)>h'(0)$ and $G_2(0)\leq1$, then $K_2(e)$ has the form \eqref{expr:K2-2}, and the optimal $e^*$ can be solved as follows: if there exists an interior point $\tilde{e}$ such that  $\tilde{e}=\arg\min_{e\in[0,e_{G_2})}K_2(e)$, $K_2(\tilde{e})\leq K_2(\hat{e})$, then $e^*=\tilde{e}$; otherwise, $e^*=\hat{e}$, where $\hat{e}$ is the minimum point of $K_2(e)$ for $e\in[e_{G_2},+\infty)$.
        \item If $G_1(0)\leq h'(0)$ and $G_2(0)\leq1$, then $K_3(e)$ has the form \eqref{expr:K3-2}, and 
        \begin{enumerate}[(a)]
            \item if there exists an interior point $\tilde{e}$ such that  $\tilde{e}={\arg\min}_{e\in(e_{G_1},e_{G_2})}K_4(e)$ and $K_4(\tilde{e})\leq \min\{K_4(\Dot{e}),K_4(\hat{e})\}$, then $e^*=\tilde{e}$;
            \item otherwise, if $K_4(\Dot{e})\leq K_4(\hat{e})$, then $e^*=\Dot{e}$;
            \item otherwise, if $K_4(\Dot{e})> K_4(\hat{e})$, then $e^*=\hat{e}$,
        \end{enumerate}
        where $\Dot{e}$ is the minimum point of $K_4(e)$ for $e\in[0,e_{G_1}]$ and $\hat{e}$ is the minimum point of $K_4(e)$ for $e\in[e_{G_2},+\infty)$.
    \end{enumerate}
\end{thm}

By substituting $e^*$ from Theorem \ref{thm:e2} into Theorem \ref{thm:alphae2}, we obtain the optimal set of insurance demand and prevention effort $(\alpha^*_{e^*},e^*)$. 
The explicit effects of premium pricing factors and risk aversion levels will be analyzed in the numerical examples provided in the next section.


\section{Numerical examples}
\label{sec:numerical}


In this section, we conduct numerical tests to illustrate the results derived in the preceding sections. 
Within the self-protection model \eqref{eq:selfprotection}, we assume that $Y$ follows a Pareto distribution with parameter $\hat{x}>0$ and $k>2$,\footnote{The reason for $k>2$ is that we need the random variable $Y$ to have finite second moment.} whose CDF is given by $F_Y(y)=1-\left(\frac{\hat{x}}{y}\right)^k$ for $y\geq\hat{x}$. 
Then, it can be calculated that 
\begin{align*}
    \mathbb{E}[X_e]&=p(e)\cdot \frac{\hat{x} k}{k-1} \quad \mbox{and}\quad\mathbb{E}[X_e^2]=p(e)\cdot \frac{\hat{x}^2 k}{k-2}.
\end{align*}
Additionally, we consider the premium principle to be a quadratic premium principle of the form $h(x)=(1+\theta_1)x+\theta_2 x^2$, where $\theta_1>0$ and $\theta_2\geq0$. 
Note that $h'(0)=1+\theta_1$. 
Furthermore, in our simulation studies, we set 
$p(e)=\frac{\gamma_1}{\gamma_2+e}$ and $c(e)=\kappa e^2$, where $0<\gamma_1<\gamma_2$ and $\kappa>0$. 
It is evident that these two functions satisfy the conditions specified in Assumptions \ref{protection:e} and \ref{cost:e}. 

\subsection{TVaR measure}
\label{sec:TVaRexample}

In this subsection, we present a numerical solution for an example of the self-protection problem under the TVaR measure, as studied in Section \ref{sec:4.1}. 
Fixing a confidence level $\beta\in(0,1)$, we then have the following: 
\begin{align*}
\text{VaR}_{\beta}(X_e)&=\begin{cases}
    0, & 0<\beta\leq 1-p(e);\\
    \text{VaR}_{\beta(e)}(Y)=F^{-1}_Y(\beta(e)), & 1-p(e)<\beta<1,
\end{cases}\\
&=\begin{cases}
    \hat{x}\left(\frac{p(e)}{1-\beta}\right)^{\frac1k}, &0\leq e<e_\beta;\\
    0, & e\geq e_{\beta}.
\end{cases}
\end{align*}
Clearly, the requirement in Assumption \ref{ass:convexTVaR} is satisfied. Consequently, we obtain 
\begin{align}\label{expr:TVaR}
    \text{TVaR}_\beta(X_e) =\frac{1}{1-\beta}\int_{\beta}^1\text{VaR}_s(X_e) \dif s=\begin{cases}
        \frac{\hat{x}k}{k-1}\cdot\left(\frac{p(e)}{1-\beta}\right)^\frac{1}{k}, & 0\leq e<e_\beta;\\
        \frac{\hat{x}k}{k-1}\cdot\frac{p(e)}{1-\beta}, & e\geq e_\beta.
    \end{cases}
\end{align}
Thus, 
$G_1(e)$ and $G_2(e)$ can be rewritten as follows:
\begin{equation}
\label{eq:TVaRG1}
    G_1(e)=\begin{cases}
        (1-\beta)^{-\frac1k}(p(e))^{\frac1k-1}, & 0\leq e<e_\beta;\\
        \frac{1}{1-\beta}, & e\geq e_\beta,
    \end{cases}
\end{equation}
and \begin{equation}
\label{eq:TVaRG2}
    G_2(e)=\begin{cases}
        \frac{k-2}{(1+\theta_1)(k-2)+2\theta_2\hat{x}(k-1)}\cdot (1-\beta)^{-\frac1k}(p(e))^{\frac1k-1}, & 0\leq e<e_\beta;\\
        \frac{k-2}{(1+\theta_1)(k-2)+2\theta_2\hat{x}(k-1)}\cdot (1-\beta)^{-1}, & e\geq e_\beta.
    \end{cases}
\end{equation}
It is clear that both objective functions (\ref{eq:TVaRG1}) and (\ref{eq:TVaRG2}) are non-decreasing on the interval $[0,e_\beta)$. 
Furthermore, $G_1$ and $G_2$ are continuous on $\mathbb{R}_+$, as $\lim_{e\rightarrow e_\beta^-}p(e)=p(e_\beta)=1-\beta$. 
It is also to note that $G_1(0)=(1-\beta)^{-\frac1k}(p(0))^{\frac1k-1}$ and $G_2(0)=\frac{k-2}{(1+\theta_1)(k-2)+2\theta_2\hat{x}(k-1)}\cdot (1-\beta)^{-\frac1k}(p(0))^{\frac1k-1}$.

\subsubsection{Optimal effort and insurance coverage w.r.t. $\theta_1$}
\label{ssec:theta1}

By setting $k=2.5$, $\hat{x}=2$, $\gamma_1=9$, $\gamma_2=25$, $\kappa=0.1$, $\beta=0.95$ and  $\theta_2=0.1$, we examine the effect of $\theta_1\in[0, 20]$ on the optimal insurance demand and the optimal protection level. 
To proceed, we first determine $\mathcal{A}_1$ and $\mathcal{A}_2$ based on the relationships among $G_1(0)$, $G_1(e_\beta)$ and $h'(0)$, 
as well as among $G_2(0)$, $G_2(e_\beta)$ and 1. 
Under the given settings, we observe that case (iii) in Theorem \ref{thm:alphae} cannot occur. 
Consequently, the subsequent numerical experiments will focus on the remaining five cases outlined in Theorem \ref{thm:alphae}. 
Let $\theta_1^{(1)}$ denote the solution of $G_2(0) = 1$, 
$\theta_1^{(2)}$ the solution of $G_1(0) = h'(0)$, $\theta_1^{(3)}$ the solution of $G_2(e_\beta) = 1$, and $\theta_1^{(4)}$  the solution of $G_1(e_\beta) = h'(0)$. 
Then we obtain $\theta_1^{(1)}=3.918$, $\theta_1^{(2)}=5.118$, $\theta_1^{(3)}=17.799$, and $\theta_1^{(4)}=19$. 

\begin{figure}[ht!]
    \centering
    \begin{minipage}[b]{0.48\textwidth}
        \centering
        \includegraphics[width=\textwidth]{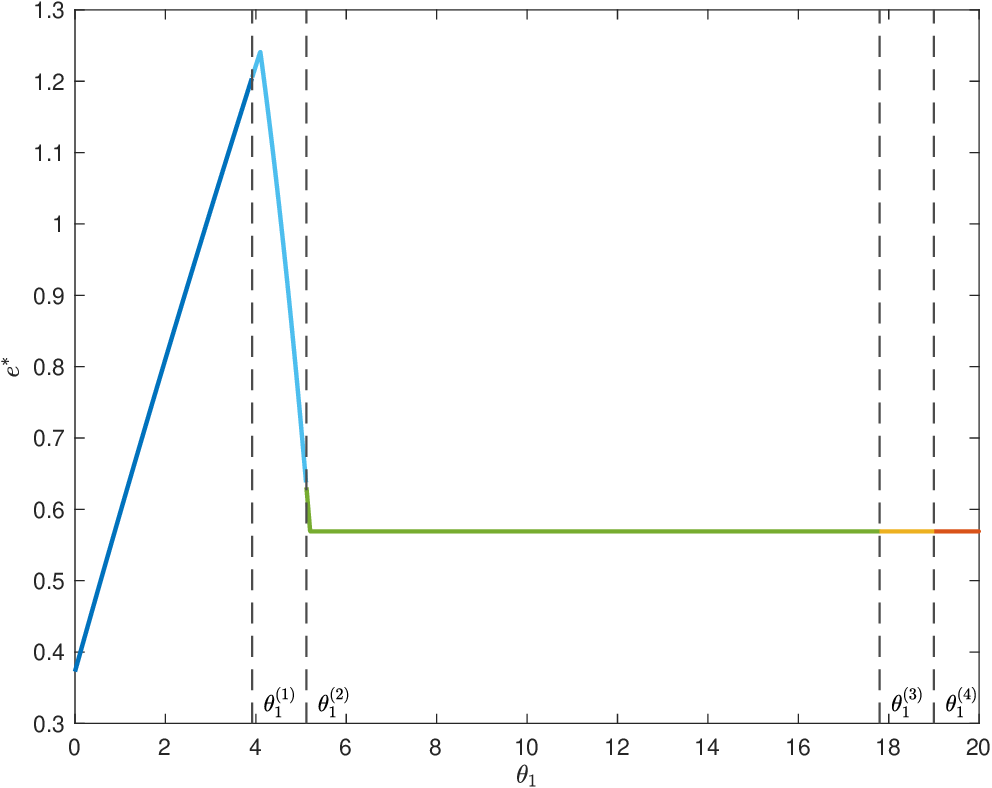}
        \caption{Optimal $e^*$ w.r.t. $\theta_1$}
        \label{fig:TVaRQ1e}
    \end{minipage}
    \hfill
    \begin{minipage}[b]{0.48\textwidth}
        \centering
        \includegraphics[width=\textwidth]{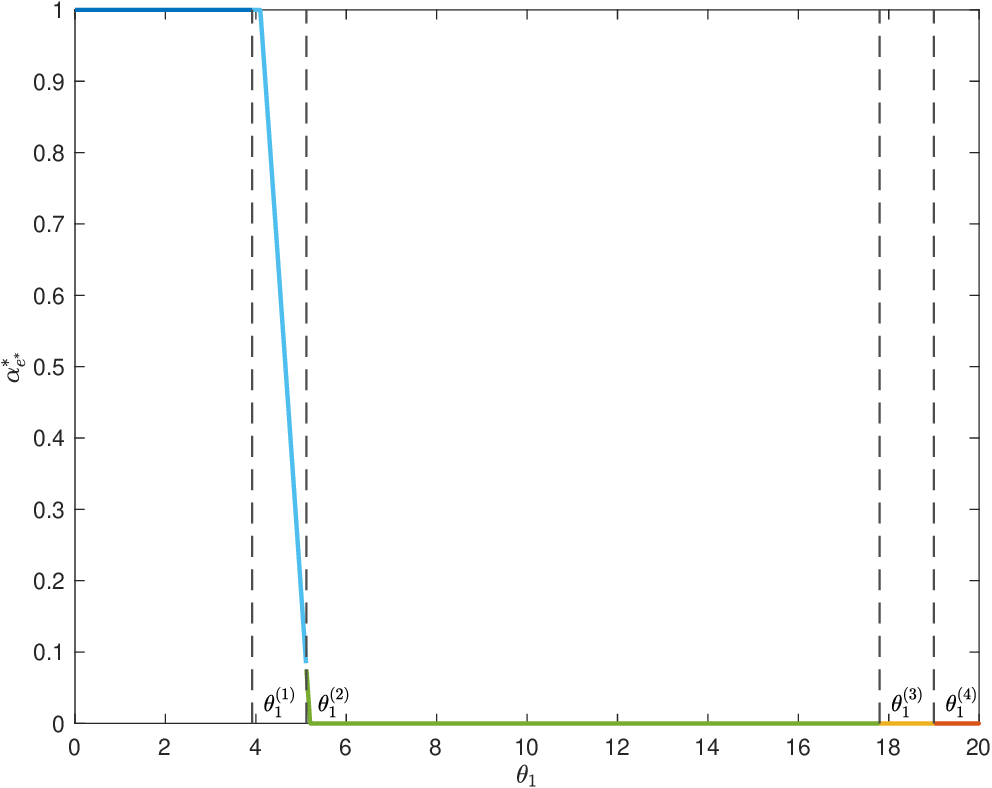}
        \caption{Optimal $\alpha_{e^*}^*$ w.r.t. $\theta_1$}
        \label{fig:TVaRQ1alpha}
    \end{minipage}
\end{figure}

Figures \ref{fig:TVaRQ1e} and \ref{fig:TVaRQ1alpha} illustrate the optimal effort $e^*$ and the optimal insurance demand $\alpha^*_{e^*}$ as the functions of the loading factor $\theta_1$, respectively. 
From Figures \ref{fig:TVaRQ1e} and \ref{fig:TVaRQ1alpha}, we observe that the values of $\theta_1^{(1)}$ to $\theta_1^{(4)}$ partition $\theta_1 \in [0,20]$ into five intervals, each corresponding to case (i), case (ii), case (iv), case (v), and case (vi) in Theorem \ref{thm:alphae} (and in Theorem \ref{thm:e}), respectively. 
From these two figures, we observe how $ \alpha_{e^*}^* $ and $ e^* $ vary with respect to the loading factor $ \theta_1 $, 
reflecting the sensitivity of insurance premiums to costs. 
Intuitively, it can be observed that the optimal effort initially increases in the interval $[0,4.1]$, then decreases in $[4.1,5.2]$, and finally stabilizes in $[5.2,20]$. 
In these three intervals, the insurance demand first corresponds to full coverage, then to partial coverage in a decreasing trend, and ultimately to no insurance.

These two figures together illustrate how individuals adjust their balance between insurance demand and self-protection in response to varying levels of $\theta_1$. 
In particular, when the insurance cost is low, individuals opt for full insurance coverage and increase their self-protection efforts as the premium loading rises, reaching a maximum at $\theta_1=4.1$. 
Beyond this point, both the insurance demand and the protection effort decrease as the premium becomes more expensive, leading to a reduced inclination for both insurance purchase and prevention measures. When the insurance cost becomes prohibitively high (i.e. $\theta_1>5.2$), individuals largely forgo insurance and stabilize at a lower level of self-protection. 
In general, the results indicate that market insurance and self-protection exhibit complementary behavior, which aligns with the classical findings of \cite{ehrlich1972market} and recent studies such as \cite{bensalem2020prevention}, \cite{bensalem2023continuous} and \cite{zeller2023risk}.


\subsubsection{Optimal effort and insurance coverage w.r.t. $\theta_2$}\label{ssec:theta2}

By setting $k=5$, $\hat{x}=1$, $\gamma_1=0.1$, $\gamma_2=0.9$, $\kappa=1$, $\beta=0.95$ and $\theta_1=5$, we study the effect of $\theta_2\in[0,20]$ on the optimal insurance demand and the optimal effort level. 
Note that $G_1(e)$ is independent of $\theta_2$ under the above settings, so we only need to consider cases (i), (ii), and (iii) in Theorem \ref{thm:alphae}. 
Let $\theta_2^{(1)}$ denote the solution of $G_2(0)=1$, and $\theta_2^{(2)}$ the solution of $G_2(e_\beta)=1$. 
It can be calculated that  $\theta_2^{(1)}=1.709$ and $\theta_2^{(2)}=5.250$.

\begin{figure}[ht!]
    \centering
    \begin{minipage}[b]{0.48\textwidth}
        \centering
        \includegraphics[width=\textwidth]{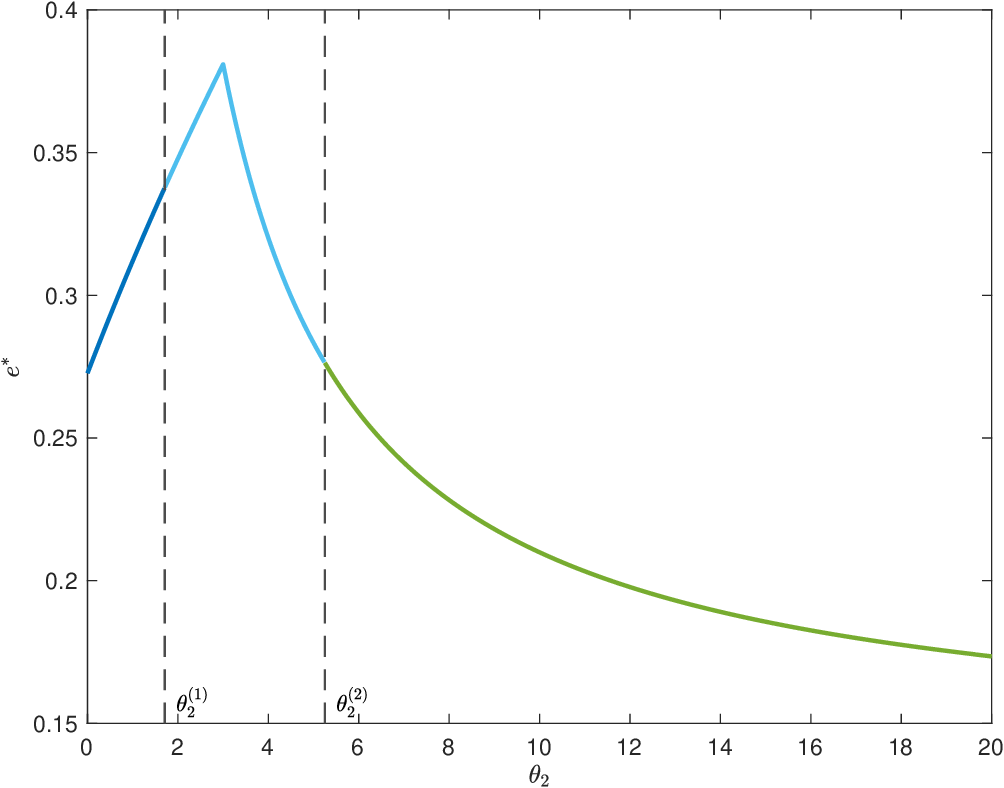}
        \caption{Optimal $e^*$ w.r.t. $\theta_2$}
        \label{fig:TVaRQ2e}
    \end{minipage}
    \hfill
    \begin{minipage}[b]{0.48\textwidth}
        \centering
        \includegraphics[width=\textwidth]{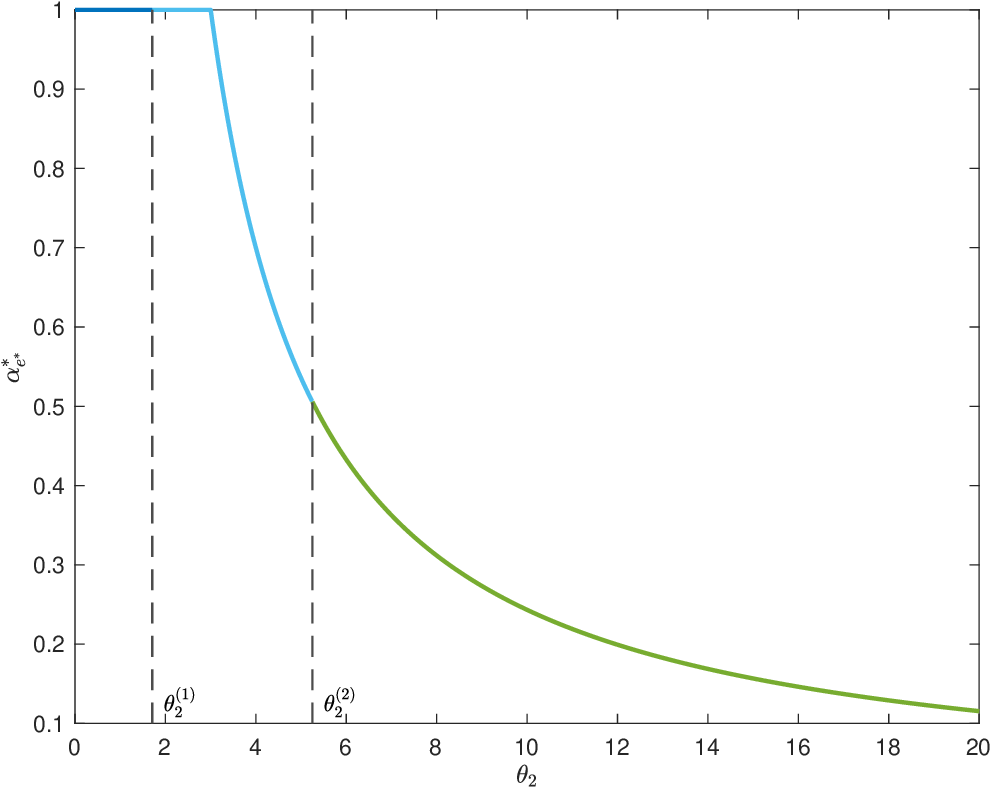}
        \caption{Optimal $\alpha_{e^*}^*$ w.r.t. $\theta_2$}
        \label{fig:TVaRQ2alpha}
    \end{minipage}
\end{figure}
Figures \ref{fig:TVaRQ2e} and \ref{fig:TVaRQ2alpha} illustrate the optimal effort $e^*$ and the optimal insurance demand $\alpha^*_{e^*}$ as the functions of the loading factor $\theta_2$, respectively. 
From Figures \ref{fig:TVaRQ2e} and \ref{fig:TVaRQ2alpha}, we observe that 
the horizontal axis is divided into three intervals by $\theta_2^{(1)}$ and $\theta_2^{(2)}$, with each interval corresponding to case (i), case (ii), and case (iii) in Theorem \ref{thm:alphae}, respectively. 
These two figures illustrate how the optimal self-protection effort $e^*$ and the insurance demand level $\alpha_{e^*}^*$ vary in response to changes in the loading factor $\theta_2$. 
Specifically, the optimal self-protection effort $e^*$ initially increases in $\theta_2\in[0,3]$ and then decreases beyond that interval. 
Simultaneously, the insurance coverage is fully purchased first and then partially demanded in a decreasing trend. 
This reinforces the complementary relationship between market insurance demand and self-protection levels as the safety loading on the second moment increases. 
This phenomenon is consistent with the findings discussed in Subsection \ref{ssec:theta1}. 
 

\subsubsection{Optimal effort and insurance coverage w.r.t. $\beta$}\label{ssec:beta}

By setting $\gamma_1 = 0.5$, $\gamma_2 = 1$, $k = 2.5$, $\hat{x} = 5$, $\theta_1 = 3$, $\theta_2 = 0.05$ and $\kappa = 0.5$, the following numerical examples focus on the effect of the risk tolerance level $\beta\in[0.5, 0.99]$ on the optimal insurance coverage and optimal effort. 
From equations \eqref{eq:TVaRG1} and \eqref{eq:TVaRG2}, we immediately know $G_1(e)$ and $G_2(e)$ are both non-increasing in $\beta\in \R_+$. 
With the above parameter settings, as $\beta$ increases, we find that case (iii) in Theorem \ref{thm:alphae} does not occur. 
Therefore, we only consider the cases in the following sequence: case (vi), case (v), case (iv), case (ii), and case (i). 
Let $\beta^{(1)}$, $\beta^{(2)}$,  $\beta^{(3)}$ and $\beta^{(4)}$ denote the solutions of $G_1(e_\beta)=h'(0)$, $G_2(e_\beta)=1$, $G_1(0)=h'(0)$ and $G_2(0)=1$, respectively. 
It can be calculated that $\beta^{(1)}=0.750$, $\beta^{(2)}=0.818$, $\beta^{(3)}=0.911$, and $\beta^{(4)}=0.960$. 
Under the given settings, these values partition the interval $(0.5,0.99)$ into five subintervals. 

\begin{figure}[ht!]
    \centering
    \begin{minipage}[b]{0.48\textwidth}
        \centering
        \includegraphics[width=\textwidth]{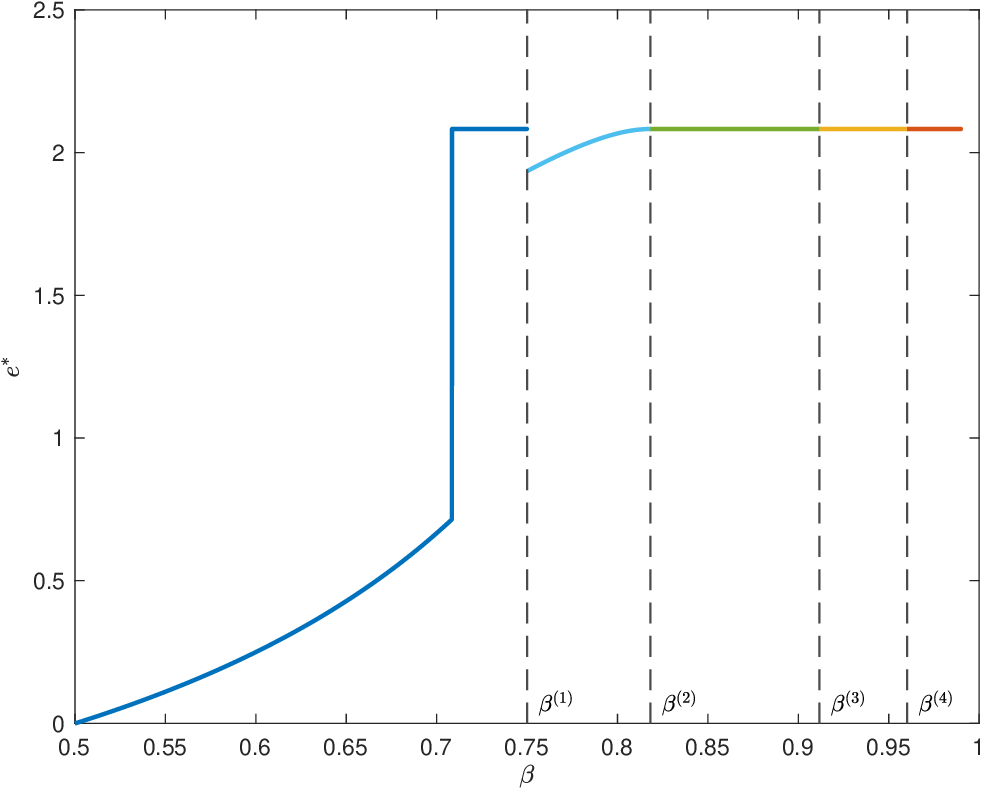}
        \caption{Optimal $e^*$ w.r.t. $\beta$}
        \label{fig:TVaRQ3e}
    \end{minipage}
    \hfill
    \begin{minipage}[b]{0.48\textwidth}
        \centering
        \includegraphics[width=\textwidth]{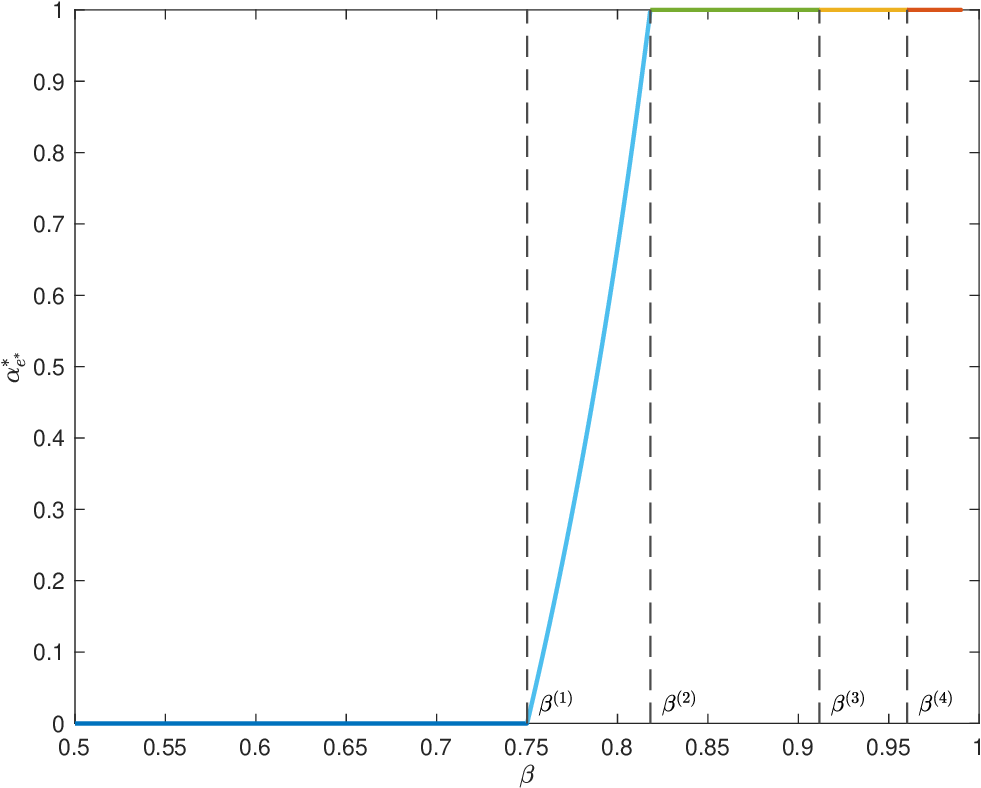}
        \caption{Optimal $\alpha_{e^*}^*$ w.r.t. $\beta$}
        \label{fig:TVaRQ3alpha}
    \end{minipage}
\end{figure}

Figures \ref{fig:TVaRQ3e} and \ref{fig:TVaRQ3alpha} depict the changes of self-protection effort $e^*$ and insurance demand $\alpha_{e^*}^*$ as $\beta$ increases. 
Note that $\beta$ can be interpreted as a parameter representing risk perception, indicating the individual's concern for extreme losses. 
As $\beta$ increases, the individual’s focus on tail risks intensifies, prompting corresponding adjustments in risk management strategies. 
For $\beta$ in $[0.5,0.75]$, the optimal effort $e^*$ increases with a sharp jump at 0.71, and then remains as a constant until the point $\beta^{(1)}=0.75$. 
Beyond this point, there is a sharp decrease in $e^*$ at $\beta^{(1)}$, followed by a slight increase before it eventually stabilizes. 
Initially, market insurance is not needed, but it is partially demanded starting from $\beta^{(1)}$, with the demand increasing until it is fully purchased beyond $\beta^{(2)}=0.818$.

In the interval $[0.5,\beta^{(1)}]$, the DM opts not to purchase any insurance. 
During this period, self-protection becomes the primary strategy for managing risk, with individuals focusing on enhancing their self-protection efforts as a proactive risk mitigation measure.
As $\beta$ approaches $\beta^{(1)}$, a noticeable shift occurs: $ e^* $ decreases sharply, and $\alpha_{e^*}^*$ begins to rise. 
This transition marks the onset of insurance demand, even as individuals scale back their self-protection efforts. 
This sharp change highlights the complementary relationship between market insurance and self-protection, where an increase in risk perception drives individuals to rely not only on self-protection but also on insurance to provide additional financial security. 
As $\beta$ continues to increase, $e^*$ stabilizes, while optimal insurance coverage gradually reaches full coverage. 
This reflects a balanced strategy where high-risk perception leads individuals to use insurance as comprehensive coverage while maintaining moderate self-protection as supplementary support. 

\subsection{Strictly convex distortion risk measure}

For the strictly concave setting, we consider the distortion function $g(u)=u^r$, $r\in(0,1)$. We then have the following $\text{VaR}_{1-u}(X_e)$ and strictly convex DRM $\rho_g(X_e)$ 
\begin{align}
    \text{VaR}_{1-u}(X_e)=\begin{cases}
        F_Y^{-1}\left(\frac{p(e)-u}{p(e)}\right)=\hat{x}\left(\frac{p(e)}{u}\right)^{1/k}, &0\leq u\leq p(e),\\
        0, &p(e)<u\leq1.
    \end{cases}
\end{align}
and
\begin{align}
    \rho_g(X_e)=\int_0^{p(e)} \text{VaR}_{1-u}(X_e) \dif g(u)=\begin{cases}
        \frac{r\hat{x}k}{kr-1}p(e)^r, &kr>1,\\
        +\infty, &\text{otherwise}.
    \end{cases}
\end{align}
Note that we are dealing with a minimization problem, and thus, infinity is not a valid solution. 
Therefore, $G_1$ and $G_2$ take the following two forms: 
\begin{equation}
    G_1(e)=\frac{r(k-1)}{kr-1}p(e)^{r-1},
\end{equation}
and
\begin{equation}
    G_2(e)=\frac{r(k-1)(k-2)}{(rk-1)[(1+\theta_1)(k-2)+2\theta_2\hat{x}(k-1)]}p(e)^{r-1}.
\end{equation}
The coefficients of $G_1$ and $G_2$ are both positive and since $r<1$, it follows that both $G_1(e)$ and $G_2(e)$ are non-decreasing in $e$. As a result, the conditions in Assumption \ref{ass:star} are fulfilled. The following examples illustrate the findings in Section \ref{sec:convexDRM}.

\subsubsection{Optimal effort and insurance coverage w.r.t. $\theta_1$}

By setting $\gamma_1 = 0.13, \gamma_2 = 10, k = 4, \theta_2 = 2, \hat{x} = 1, r = 0.5, \kappa = 0.01$, we examine the effect of $\theta_1\in[0, 20]$ on the optimal insurance demand and the optimal effort level. 
Let $\theta_1^{(1)}$ and $\theta_1^{(2)}$ denote the solutions of equations $G_2(0)=1$ and $G_1(0)=h'(0)$, respectively. 
We find that $\theta_1^{(1)}=6.155$ and $\theta_1^{(2)}=12.155$. 
Thus, $\theta_1^{(1)}$ and $\theta_1^{(2)}$ partition the interval $[0,20]$ into three regions, corresponding to case (i), case (ii), and case (iii) in Theorem \ref{thm:alphae2}. 

\begin{figure}[ht!]
    \centering
    \begin{minipage}[b]{0.48\textwidth}
        \centering
        \includegraphics[width=\textwidth]{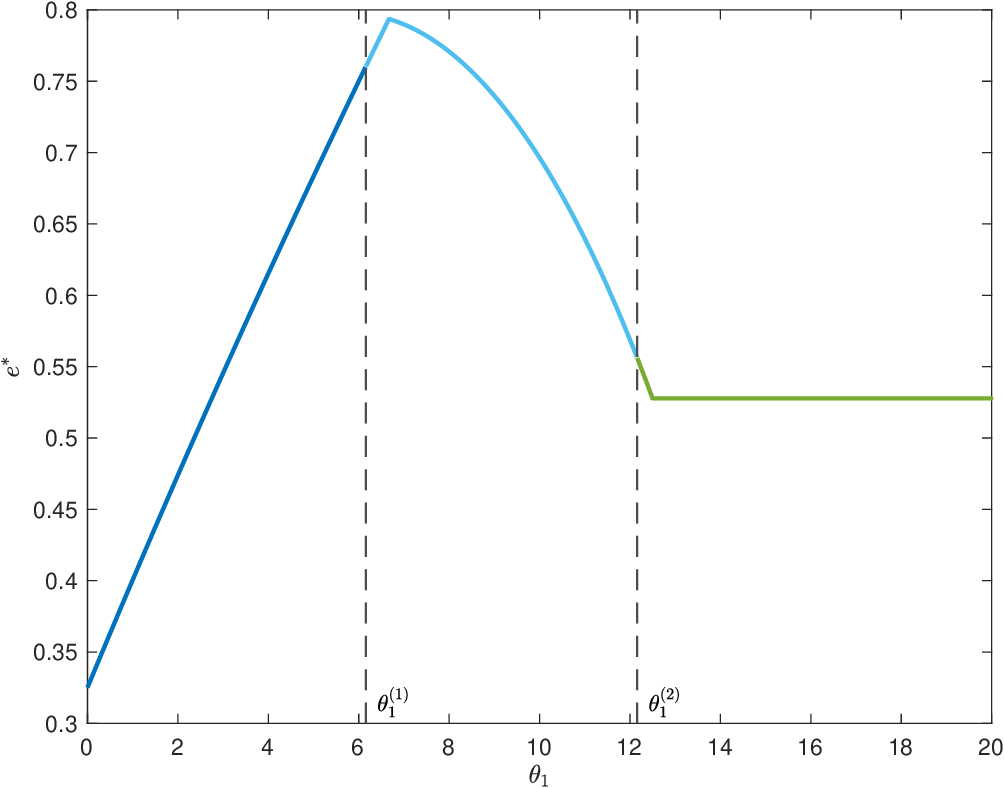}
        \caption{Optimal $e^*$ w.r.t. $\theta_1$}
        \label{fig:DRMQ1e}
    \end{minipage}
    \hfill
    \begin{minipage}[b]{0.48\textwidth}
        \centering
        \includegraphics[width=\textwidth]{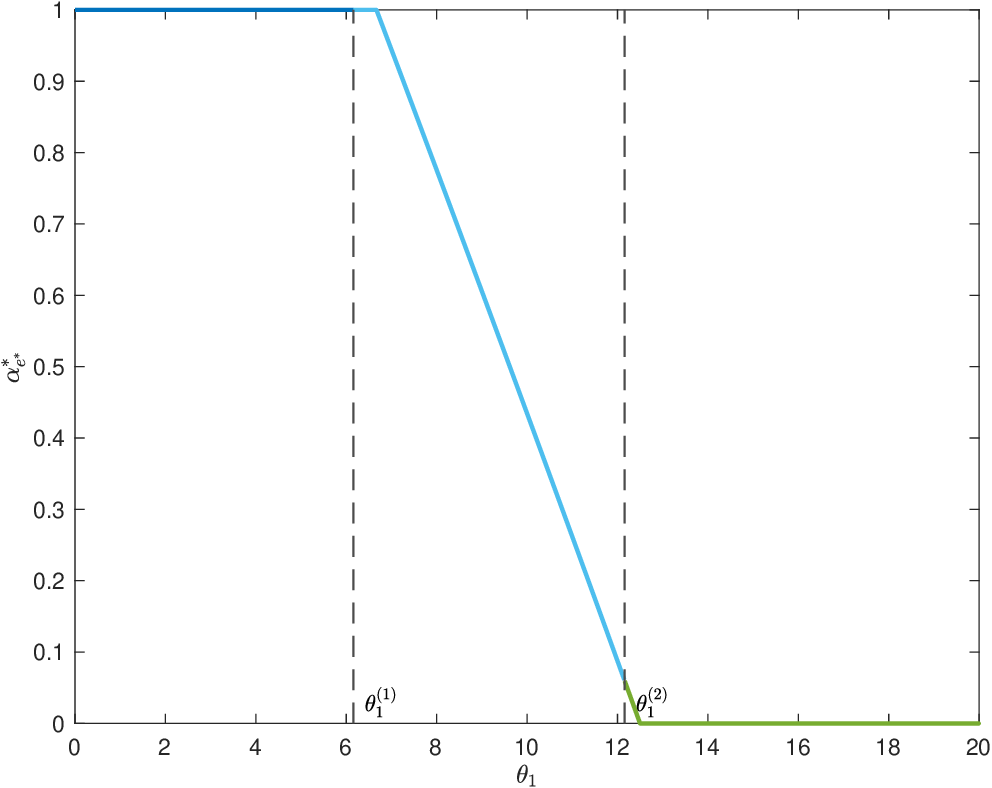}
        \caption{Optimal $\alpha_{e^*}^*$ w.r.t. $\theta_1$}
        \label{fig:DRMQ1alpha}
    \end{minipage}
\end{figure}

Figures \ref{fig:DRMQ1e} and \ref{fig:DRMQ1alpha} depict the optimal effort $e^*$ and optimal insurance demand $\alpha_{e^*}^*$ as functions of the parameter $\theta_1$, respectively. 
From these figures, we can see that as $\theta_1$ increases, $e^*$ rises initially. 
It reaches its peak around $\theta_1\approx 6.8$, after which $e^*$ decreases and stabilizes. 
Meanwhile, $\alpha_{h,e}^*$ starts at 1, then gradually decreases, ultimately reaching 0. 
This inverse relationship between insurance demand and self-protection confirms the complementary effect between market insurance and self-protection. 
This finding aligns with the results presented in Subsection \ref{ssec:theta1}.

\subsubsection{Optimal effort and insurance coverage w.r.t. $\theta_2$}

By setting $\gamma_1 = 0.007, \gamma_2 = 1, k = 3, \theta_1 = 2, \hat{x} = 1, r = 0.5, \kappa = 0.01$, we examine the effect of $\theta_2\in[0, 20]$ on the optimal insurance demand and the optimal protection level. 
Note that $G_1$ is independent of $\theta_2$ and $G_1(0)>h'(0)$. 
As a result, case (iii) in Theorem \ref{thm:alphae2} does not occur. 
Consequently, we will focus on case (i) and case (ii) in Theorem \ref{thm:alphae2} in the following discussions. 
Let $\theta_2^{(1)}$ denote the solution of equation $G_2(0)=1$. 
Then it divides the interval $[0, 20]$ into two parts, with the left side corresponding to case (i) and the right side to case (ii). 

\begin{figure}[ht!]
    \centering
    \begin{minipage}[b]{0.48\textwidth}
        \centering
        \includegraphics[width=\textwidth]{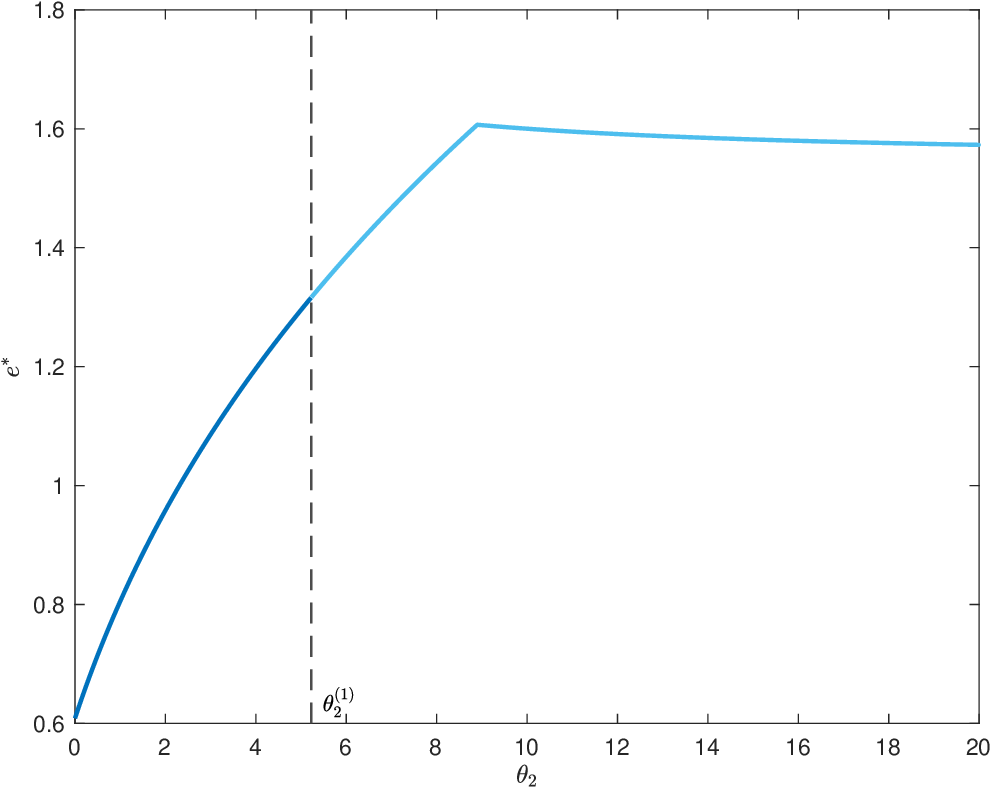}
        \caption{Optimal $e^*$ w.r.t. $\theta_2$}
        \label{fig:DRMQ2e}
    \end{minipage}
    \hfill
    \begin{minipage}[b]{0.48\textwidth}
        \centering
        \includegraphics[width=\textwidth]{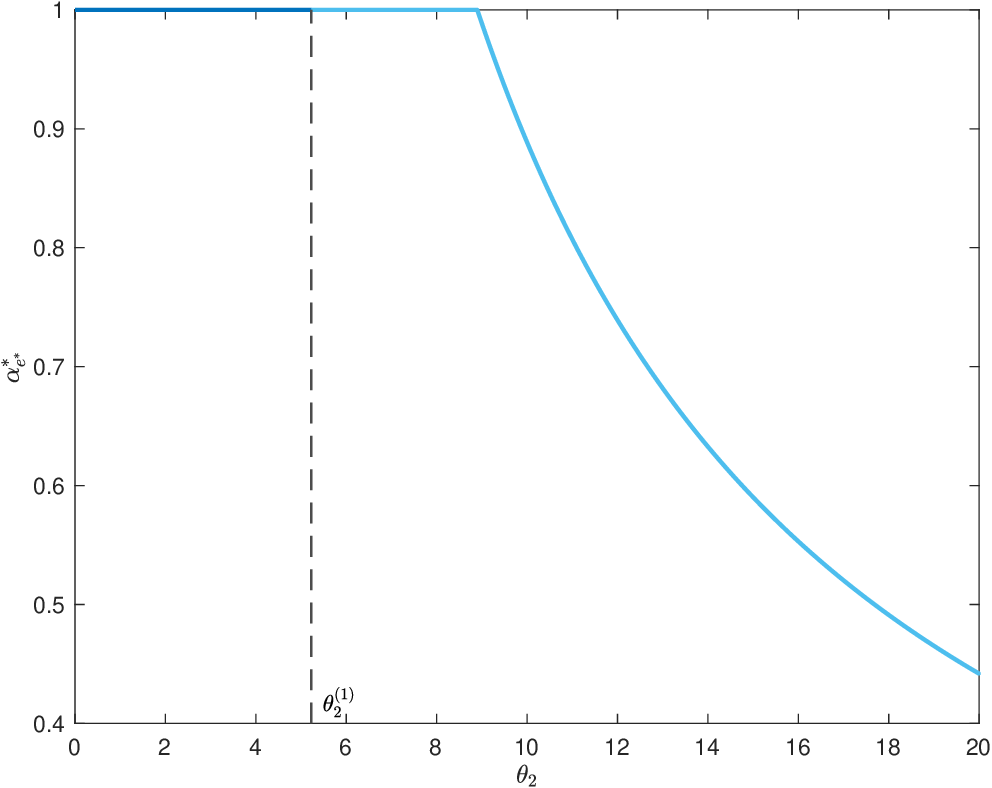}
        \caption{Optimal $\alpha_{e^*}^*$ w.r.t. $\theta_2$}
        \label{fig:DRMQ2alpha}
    \end{minipage}
\end{figure}

Figures \ref{fig:DRMQ2e} and \ref{fig:DRMQ2alpha} depict the optimal effort $e^*$ and the optimal insurance demand $\alpha_{e^*}^*$ as the function of the parameter $\theta_2$, respectively. 
It is evident that the complementary effect between insurance demand and self-protection is strongly supported by the increasing premium loading on the quadratic term. 
This finding is consistent with the analysis presented in Subsection \ref{ssec:theta2}. 

\subsubsection{Optimal effort and insurance coverage w.r.t. $r$}

By setting $\gamma_1 = 0.02, \gamma_2 = 1, k = 4, \theta_1 = 1, \theta_2 = 2, \hat{x} = 1, \kappa = 0.01$, we explore the impact of the risk aversion degree $r\in[0.5,0.95]$ on optimal effort and insurance coverage. 
Let $r^{(1)}$ and $r^{(2)}$ denote the optimal solutions to the equations $G_1(0)=h'(0)$ and $G_2(0)=1$, respectively. 
Then they partition the interval $[0.5,0.95]$ into three regions, corresponding to the case (i), case (ii), and case (iii) in Theorem \ref{thm:alphae2}. 

\begin{figure}[ht!]
    \centering
    \begin{minipage}[b]{0.48\textwidth}
        \centering
        \includegraphics[width=\textwidth]{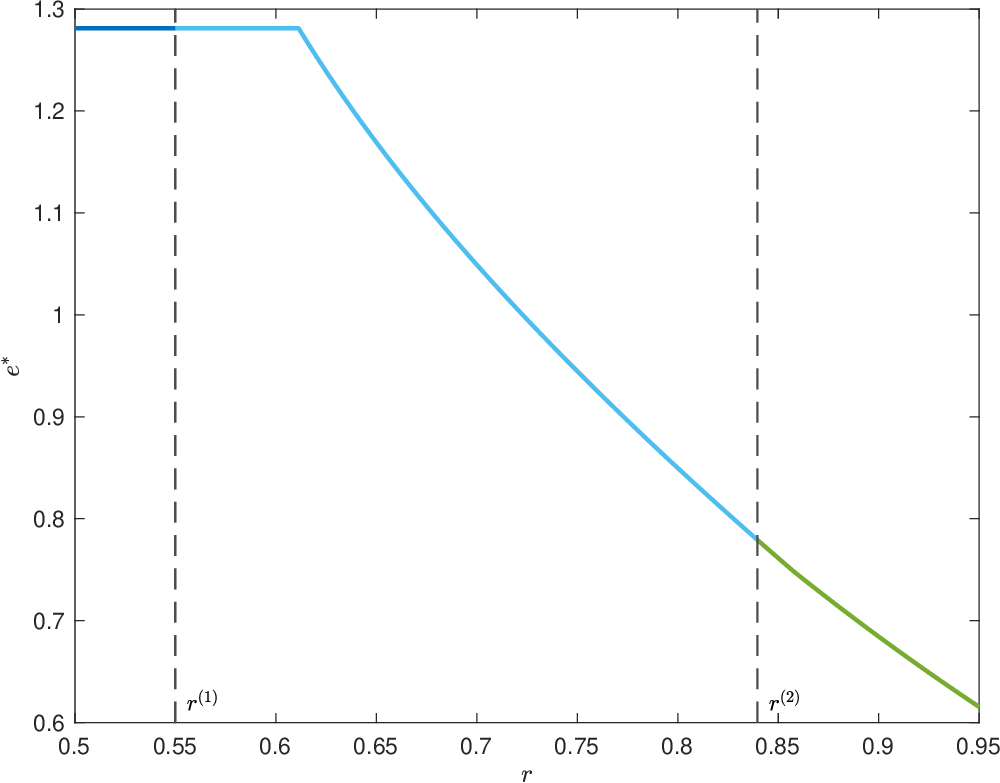}
        \caption{Optimal $e^*$ w.r.t. $r$}
        \label{fig:DRMQ3e}
    \end{minipage}
    \hfill
    \begin{minipage}[b]{0.48\textwidth}
        \centering
        \includegraphics[width=\textwidth]{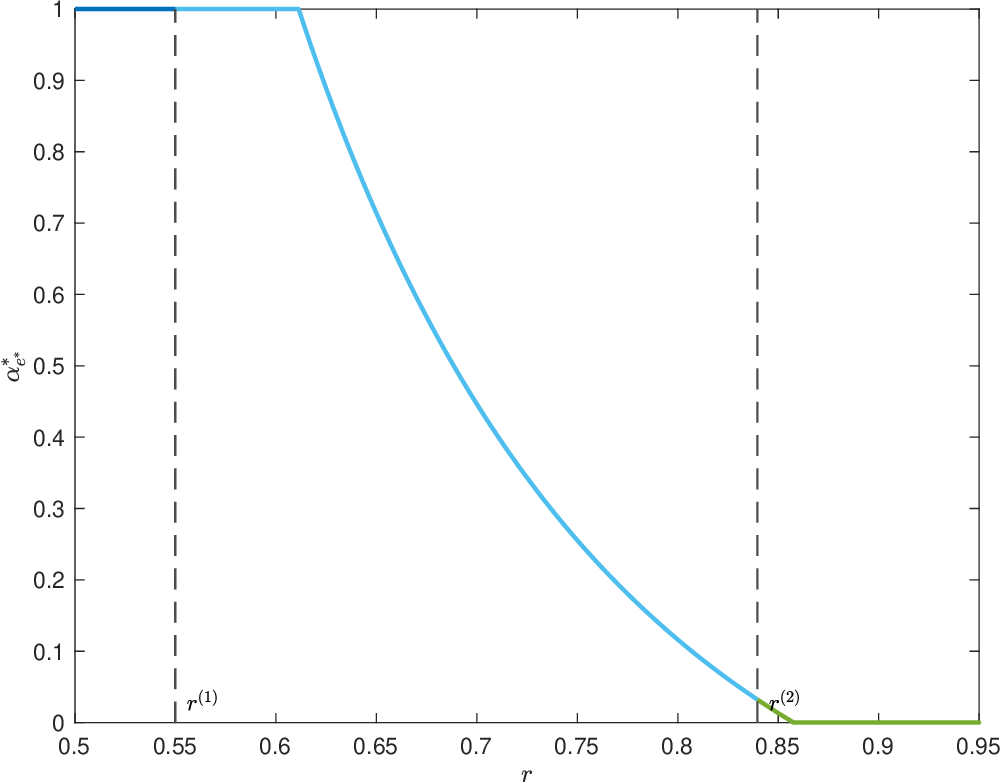}
        \caption{Optimal $\alpha_{e^*}^*$ w.r.t. $r$}
        \label{fig:DRMQ3alpha}
    \end{minipage}
\end{figure}

Figures \ref{fig:DRMQ3e} and \ref{fig:DRMQ3alpha} depict the optimal effort $e^*$ and the optimal insurance demand $\alpha_{e^*}^*$ as the functions of the parameter $r$ in strictly concave distortion function $g(u)=u^r$. 
As observed from these figures, an increase in the degree of risk aversion (i.e., a decrease in the value of $r$) leads to higher demand for both market insurance and the self-protection level. 
Notably, the pattern for the optimal protection level changes continuously with respect to the value of $r$, in stark contrast to the discontinuous behavior observed under the TVaR setting discussed in Subsection \ref{ssec:beta}.

\section{Effects of ex ante moral hazard}\label{sec:moral}

In numerous practical insurance scenarios, the precise level of protection effort exerted by the insured may not be observable by the insurer, resulting in ex ante moral hazard that influences the DM's optimal strategy. 
This issue has been extensively discussed and investigated in the literature, see, for example, \cite{seog2012moral}, \cite{winter2013optimal}, \cite{seog2024moral}, and references therein. 
Under the presence of ex ante moral hazard, the DM seeks to address the following optimization problem: 
\begin{equation}
\label{prob:moral}
\begin{aligned}
\min_{(e,\alpha)\in\mathbb{R}_{+}\times[0,1]} & \quad \rho_{g}\left(X_e-\alpha X_e+\mathbb{E}[h(\alpha X_e)]+c(e)\right) \\
\mbox{s.t.}\quad\;\;  &  \quad e=\arg\min_{b\in\mathbb{R}_{+}}\rho_{g}\left(X_b-\alpha X_b+\mathbb{E}[h(\alpha X_e)]+c(b)\right),
\end{aligned}
\end{equation} 
where the constraint is widely known as \emph{incentive compatibility} conditions in the literature, see, e.g., \cite{winter2013optimal}. 
Denote the solution of problem (\ref{prob:moral}) by $(\ddot{e}^*,\ddot{\alpha}^*_{\ddot{e}^*})$. 
By using the translation invariance and positive homogeneity of DRMs, it directly follows that 
$\rho_g(X_b-\alpha X_b +\mathbb{E}[h(\alpha X_e)]+c(b))=(1-\alpha)\rho_g(X_b)+c(b)+\mathbb{E}[h(\alpha X_e)]$. 
Let
\begin{equation}
    f(b):=
    (1-\alpha)\rho_{g}(X_b)+c(b),\quad b\geq0.
\end{equation}
Then the constraint of problem (\ref{prob:moral}) is equivalent to $e=\arg\min_{b\in\mathbb{R}_{+}} f(b)$. 

If $\alpha=1$, then $f(b)=c(b)$ and consequently, the optimal solution to minimize $c(b)$ is uniquely achieved at $b=0$ since the function $c(\cdot)$ satisfies Assumption \ref{cost:e}. 
For this case, a corner solution to problem (\ref{prob:moral}) is achieved at $(\ddot{e}^*_{\ddot{\alpha}^*},\ddot{\alpha}^*)=(0,1)$. 

Now, we focus on the non-trivial case that $0\leq\alpha<1$. 
%
%
In the following, we assume $g$ is concave. 
Under Assumptions \ref{cost:e} and \ref{ass:convexTVaR}, it can be easily checked that $f(b)$ is strictly convex in $b\in\mathbb{R}_+$ as the nonnegative weighted sums of a strictly convex function ($c(e)$) and a convex function ($\rho_g(X_e)$). 
Moreover, for any $b\in \R_+$, we have 
\begin{equation*}
     f'(b)= (1-\alpha) \rho_g'(X_b) + c'(b), 
\end{equation*}
where $\rho_g'(X_b):=\frac{\dif}{\dif e} \rho_g(X_e)|_{e=b}$. 
Note also that $c(e)$ satisfies Assumption \ref{cost:e}, then we have that $f'(b)$
satisfies 
\[
f'(0)=(1-\alpha)\rho'_{g}(X_0)<0\quad\mbox{and}\quad f'(\infty)=\infty. 
\]
Hence, the optimal protection effort $e$ and insurance demand $\alpha$ should satisfy 
\begin{equation*}
    (1-\alpha)\rho_{g}'(X_e)+c'(e)=0,
\end{equation*}
which yields that
\begin{equation}\label{equa:alphaestar}
    \alpha=1+\frac{c'(e)}{\rho_{g}'(X_e)},\quad \alpha\in[0,1).
\end{equation}

With the help of \eqref{equa:alphaestar} and Assumption \ref{cost:e}, 
the admissible set of protection effort can be denoted as
\begin{equation*}
    \mathcal{B}=\left\{e\big{|} -1\leq\frac{c'(e)}{\rho_{g}'(X_e)}<0\right\}=\left\{e|e>0,~ c'(e)+\rho_{g}'(X_e)\leq0\right\}.
\end{equation*}
The economical interpretation of the effort set $\mathcal{B}$ is understood as follows: the DM would like to execute non-zero effort such that the marginal risk position reduction per unit of effort should be strictly larger than the marginal cost. 
Moreover, from Assumption 
\ref{cost:e}, 
the function $c'(e)+\rho_{g}'(X_e)$ is strictly increasing and continuous in $e$. 
Since $c'(0)=0$, $c'(\infty)=\infty$ and $\rho_{g}'(X_0)<0$, it then follows that  $c'(e)+\rho_{g}'(X_e)=0$ has a unique solution, denoted as $e_{\mathcal{B}}$. Hence, $\mathcal{B}=(0,e_{\mathcal{B}}]$.

Let $L(e)$ be the objective function of (\ref{prob:moral}) subject to the moral hazard constraint. 
It follows that
\begin{equation}\label{expr:L_temp}
   L(e)=\begin{cases}
        \mathbb{E}[h(X_{0})], &e=0;\\
        -\frac{\rho_{g}(X_e)c'(e)}{\rho_{g}'(X_e)}+\mathbb{E}\left[h\left(X_e\left(1+\frac{c'(e)}{\rho_{g}'(X_e)}\right)\right)\right]+c(e), &e\in\mathcal{B}.
    \end{cases}
\end{equation}

For ease of expression, we will denote $H(e) = 1 + \frac{c'(e)}{\rho_g'(X_e)}$ from now on, and from equation \eqref{equa:alphaestar} we have $\alpha=H(e)$. 
Consequently, the equation above can be further rewritten as:
\begin{equation}\label{expr:L}
    L(e)=\begin{cases}
        \mathbb{E}[h(X_{0})], &e=0;\\
        \rho_{g}(X_e)(1-H(e)) + \mathbb{E}\left[h\left(H(e) X_e\right)\right]+c(e), &e\in\mathcal{B}.
    \end{cases}
\end{equation} 

For simplicity, we focus on the TVaR measure throughout this section. 
For TVaR measure, it follows from  \eqref{TVaR:express} that
\begin{equation}
    \frac{\dif}{\dif e}\text{TVaR}_\beta(X_e)=\begin{cases}
        \frac{p'(e)}{1-\beta}\int_{\beta(e)}^1 \text{VaR}_t(Y)\dif t-\frac{p'(e)}{p(e)}\text{VaR}_{\beta(e)}(Y), &0\leq e<e_\beta;\\
        \frac{\mathbb{E}[Y]}{1-\beta}p'(e), &e\geq e_\beta.
    \end{cases}
\end{equation}
Consequently, we can directly obtain the explicit expression for $H(e)$: 
\begin{equation}
    H(e)=\begin{cases}
        1+c'(e)\cdot \left(\frac{p'(e)}{1-\beta}\int_{\beta(e)}^1 \text{VaR}_t(Y)\dif t-\frac{p'(e)}{p(e)}\text{VaR}_{\beta(e)}(Y)\right)^{-1}, & 0\leq e<e_\beta;\\
        1+\frac{(1-\beta)c'(e)}{\mathbb{E}[Y]p'(e)}, &e\geq e_\beta.  
    \end{cases}
\end{equation}
Let $H_1(e) = 1+c'(e)\cdot \left(\frac{p'(e)}{1-\beta}\int_{\beta(e)}^1 \text{VaR}_t(Y)\dif t-\frac{p'(e)}{p(e)}\text{VaR}_{\beta(e)}(Y)\right)^{-1}$ and $H_2(e)=1+\frac{(1-\beta)c'(e)}{\mathbb{E}[Y]p'(e)}$. 
Then 
\[
H(e)=H_1(e)\mathds{1}_{e\in[0,e_\beta)}+H_2(e)\mathds{1}_{e\in[e_\beta,+\infty)}.
\]
Therefore, from equation \eqref{expr:L}, the objective function of problem (\ref{prob:moral}) subject to the moral hazard constraint is 
\begin{equation}
\label{expr:LTVaR}
   L(e)=\begin{cases}
        \mathbb{E}[h(X_{0})], &e=0;\\
        \text{TVaR}_\beta(X_e)(1-H_1(e)) + \mathbb{E}\left[h\left(H_1(e) X_e\right)\right]+c(e), &e\in\mathcal{B}\cap(0,e_{\beta});\\
        \text{TVaR}_\beta(X_e)(1-H_2(e)) + \mathbb{E}\left[h\left(H_2(e) X_e\right)\right]+c(e), &e\in\mathcal{B}\cap[e_{\beta},\infty).
    \end{cases}
\end{equation}

In the following, we present numerical examples to illustrate the impact of moral hazard on optimal effort and optimal insurance coverage. 
The parameter settings in the following subsections are consistent with those outlined in subsections \ref{ssec:theta1}, \ref{ssec:theta2}, and \ref{ssec:beta}, respectively. 
This setting allows for a comprehensive examination of the effects of parameters $\theta_1,\theta_2$ and $\beta$ on optimal effort and coverage, while also facilitating a comparison with the conclusions drawn in the earlier sections. 
From equation \eqref{expr:TVaR}, we have 
\[
H_1(e) = 1-\frac{2\kappa(k-1)}{\hat{x}}(1-\beta)^{\frac1k}\gamma_1^{-\frac1k} e (\gamma_2+e)^{\frac1k+1}
\] 
and 
\[
H_2(e) = 1-\frac{2\kappa(k-1)}{\hat{x}k\gamma_1}(1-\beta)e(\gamma_2+e)^2. 
\]

\subsection{Optimal effort and insurance coverage w.r.t. $\theta_1$}
\label{ssec:MHtheta1}

By setting $k=2.5$, $\hat{x}=2$, $\gamma_1=9$, $\gamma_2=25$, $\kappa=0.1$, $\beta=0.95$ and  $\theta_2=0.1$, 
we examine the effect of $\theta_1\in[0, 20]$ on the optimal insurance demand and the optimal protection level. 
Note that we only consider values of $e$ for which the range of $H(e)$ lies within $[0,1)$. 
Under these numerical settings, the function $H(e) $ is discontinuous at $ e_\beta $, but it remains a continuously non-increasing function on both intervals. 
Through calculation, we find that $ H(e_\beta) < 0 $ and $ H(0) H(e_\beta-0) < 0 $. Therefore, there must exist a zero point $ e_{\mathcal{B}} $ in the interval $(0,e_\beta)$. 
After calculation, we obtain $e_{\mathcal{B}}=0.569135$. 
From \eqref{expr:LTVaR}, we know that the objective function with moral hazard constraint can be simplified into the following form,
\begin{equation}\label{expr:Q1Le}
    L(e)= \begin{cases}
        \frac{(1+\theta_1)\hat{x}k}{k-1}p(0)+\frac{\theta_2\hat{x}^2k}{k-2}, &e=0;\\
        \frac{\hat{x}k}{(k-1)(1-\beta)^{\frac1k}}p(e)^{\frac1k}(1-H_1(e))+ \frac{(1+\theta_1)\hat{x}k}{k-1}H_1(e)p(e)+\frac{\theta_2\hat{x}^2k}{k-2} H_1^2(e)p(e)+c(e), & 0<e\leq e_{\mathcal{B}}.
    \end{cases}
\end{equation}
\begin{figure}[ht!]
    \centering
    \begin{minipage}[b]{0.48\textwidth}
        \centering
        \includegraphics[width=\textwidth]{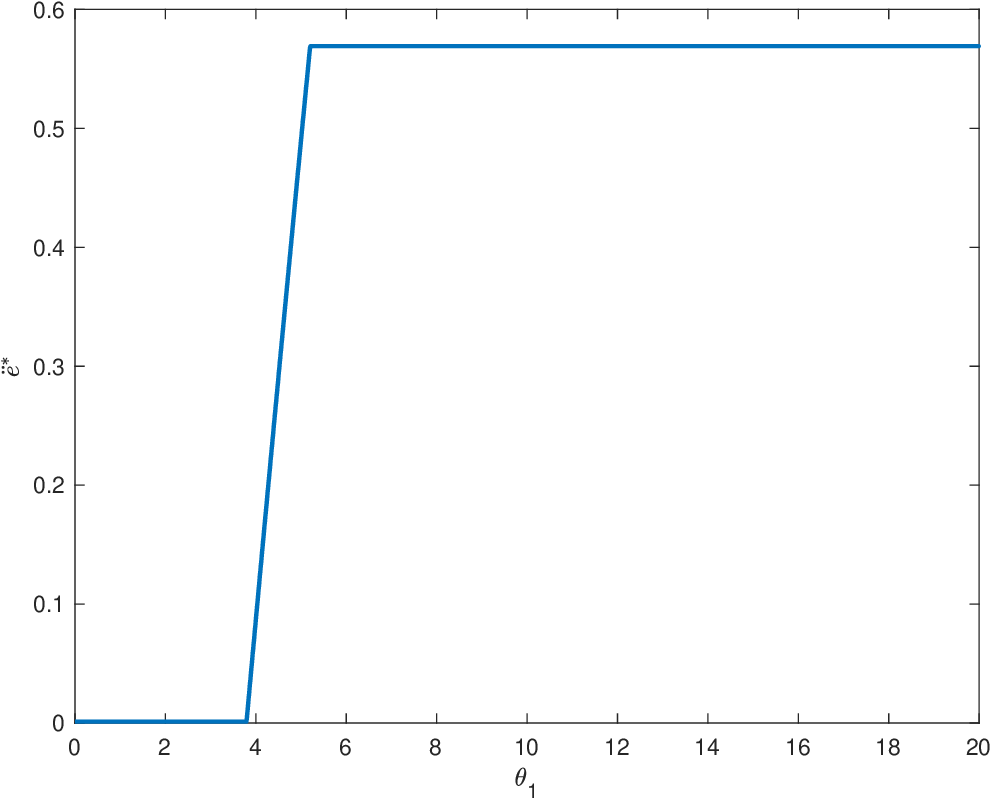}
        \caption{Optimal $\ddot{e}^*$ w.r.t. $\theta_1$}
        \label{fig:MHQ1e}
    \end{minipage}
    \hfill
    \begin{minipage}[b]{0.48\textwidth}
        \centering
        \includegraphics[width=\textwidth]{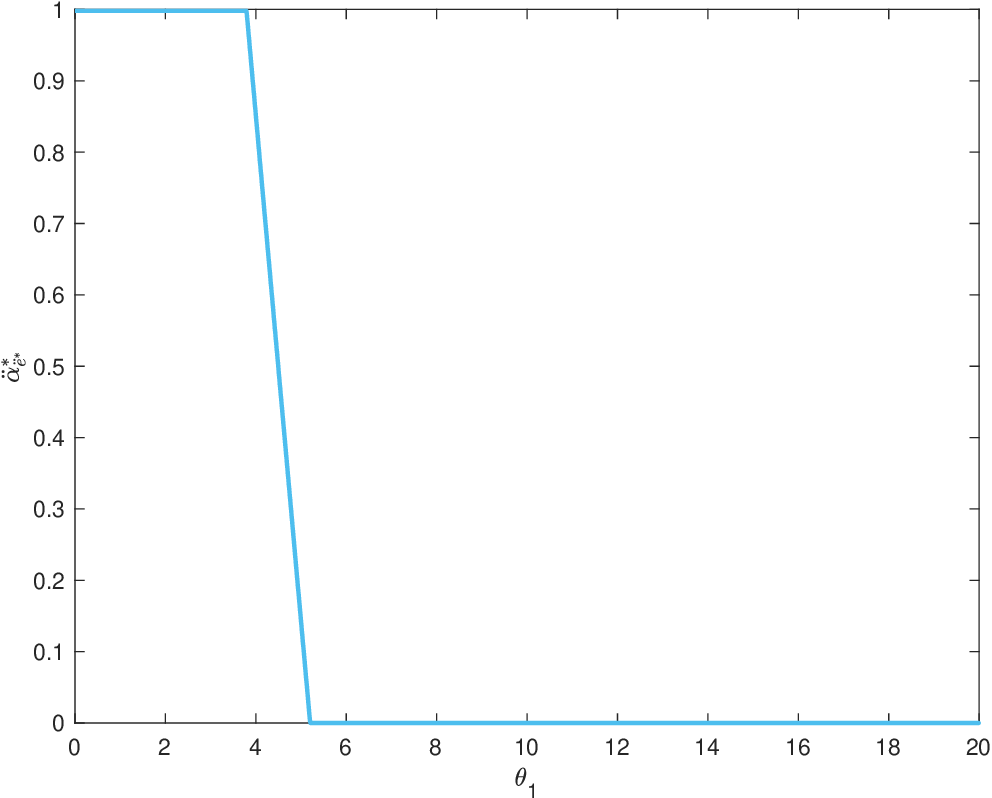}
        \caption{Optimal $\ddot{\alpha}_{\ddot{e}^*}^*$ w.r.t. $\theta_1$}
        \label{fig:MHQ1alpha}
    \end{minipage}
\end{figure}

Figures \ref{fig:MHQ1e} and  \ref{fig:MHQ1alpha} illustrate the relationship between insurance price $\theta_1$ and optimal effort $\ddot{e}^*$ and optimal insurance coverage $\ddot{\alpha}_{\ddot{e}^*}^*$, respectively. 
From these figures, we observe that as $\theta_1$ increases, individuals gradually shift from relying on insurance to relying more on effort, demonstrating the substitutability between insurance and self-protection efforts. 
This finding contrasts with the results in Section \ref{ssec:theta1}, where the relationship between insurance and effort follows a different pattern.

\subsection{Optimal effort and insurance coverage w.r.t. $\theta_2$}

By setting $k=5$, $\hat{x}=1$, $\gamma_1=0.1$, $\gamma_2=0.9$, $\kappa=1$, $\beta=0.95$ and $\theta_1=5$, we study the influence of $\theta_2\in[0,20]$ on the optimal solution for the DM. 
It is easy to see that $H(e)$ is independent of $\theta_2$. 
Like subsection \ref{ssec:MHtheta1}, under these parameters, $e_\mathcal{B}=0.569135\in (0,e_\beta)$. Therefore, our objective function is the same as \eqref{expr:Q1Le}. 
\begin{figure}[ht!]
    \centering
    \begin{minipage}[b]{0.48\textwidth}
        \centering
        \includegraphics[width=\textwidth]{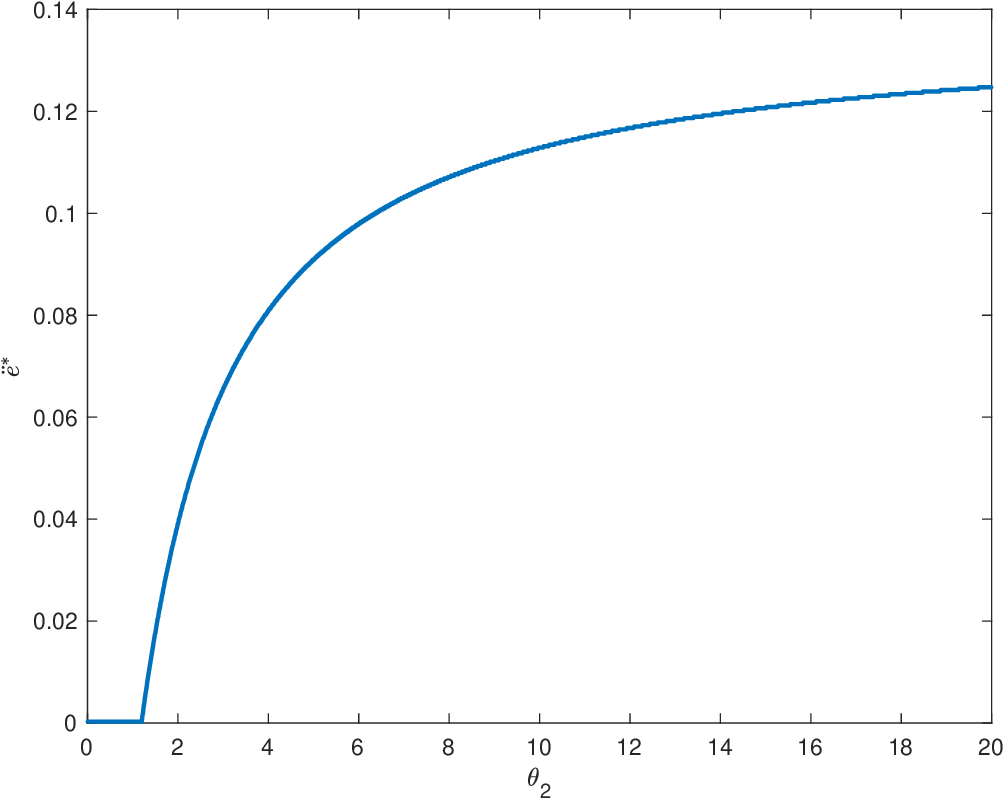}
        \caption{Optimal $\ddot{e}^*$ w.r.t. $\theta_2$}
        \label{fig:MHQ2e}
    \end{minipage}
    \hfill
    \begin{minipage}[b]{0.48\textwidth}
        \centering
        \includegraphics[width=\textwidth]{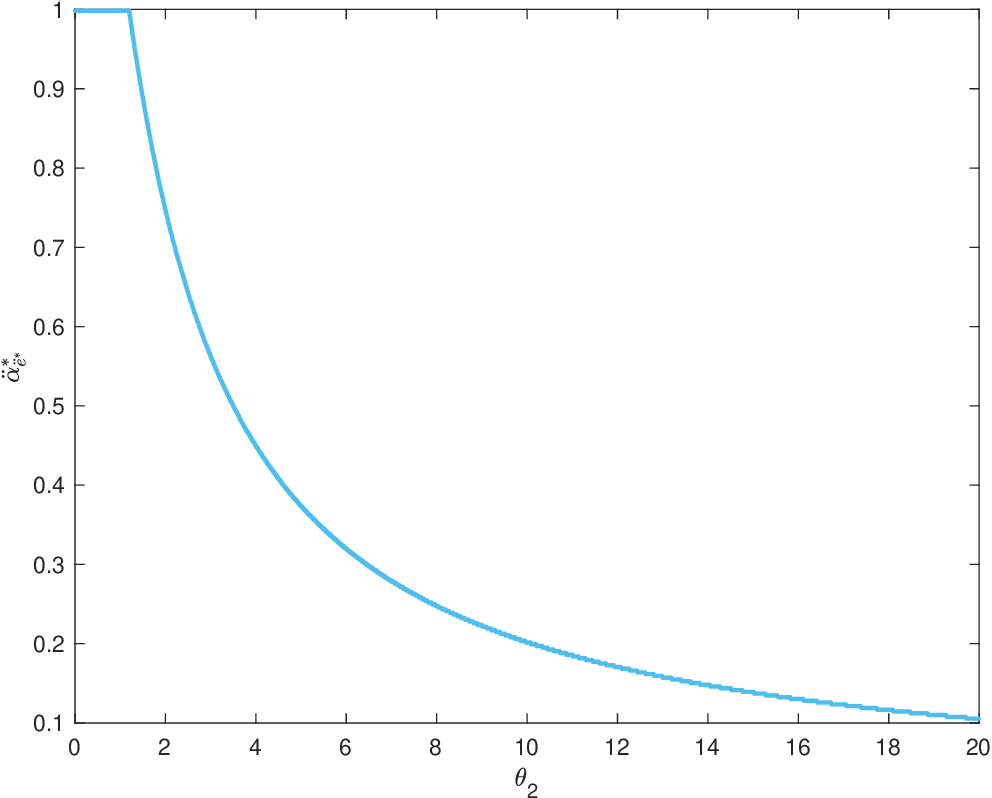}
        \caption{Optimal $\ddot{\alpha}_{\ddot{e}^*}^*$ w.r.t. $\theta_2$}
        \label{fig:MHQ2alpha}
    \end{minipage}
\end{figure}

Figures \ref{fig:MHQ2e} and \ref{fig:MHQ2alpha} illustrate the relationship between the premium $\theta_2$ and the optimal insurance coverage as well as the optimal effort. 
The results shown in these two figures are consistent with those in Section \ref{ssec:MHtheta1}, reflecting the substitutability between insurance and effort.

\subsection{Optimal effort and insurance coverage w.r.t. $\beta$}

By setting $\gamma_1 = 0.5$, $\gamma_2 = 1$, $k = 2.5$, $\hat{x} = 5$, $\theta_1 = 3$, $\theta_2 = 0.05$ and $\kappa = 0.5$, the following numerical examples exhibit the effect of the risk tolerance level $\beta\in[0.5, 0.99]$ on the optimal solution. 
Since $H(e)$ is dependent of $\beta$, the relationship between $e_{\mathcal{B}}$ and $e_\beta$ changes with the value of $\beta$. 
When $\beta$ is relatively small, $e_\beta$ is less than $e_{\mathcal{B}}$, so we have \eqref{expr:Q3Le}. 
However, when $\beta$ is relatively large, $e_\beta$ exceeds $e_{\mathcal{B}}$, and we have \eqref{expr:Q1Le}.

\begin{equation}
\label{expr:Q3Le}
    L(e)= \begin{cases}
        \frac{(1+\theta_1)\hat{x}k}{k-1}p(0)+\frac{\theta_2\hat{x}^2k}{k-2}, &e=0;\\
        \frac{\hat{x}k}{(k-1)(1-\beta)^{\frac1k}}p(e)^{\frac1k}(1-H_1(e))+ \frac{(1+\theta_1)\hat{x}k}{k-1}H_1(e)p(e)+\frac{\theta_2\hat{x}^2k}{k-2} H_1^2(e)p(e)+c(e), & 0 < e < e_\beta;\\
        \frac{\hat{x}k}{(k-1)(1-\beta)^{\frac1k}}p(e)^{\frac1k}(1-H_2(e))+ \frac{(1+\theta_1)\hat{x}k}{k-1}H_2(e)p(e)+\frac{\theta_2\hat{x}^2k}{k-2} H_2^2(e)p(e)+c(e), & e_\beta\leq e\leq e_{\mathcal{B}}.
    \end{cases}
\end{equation}
\begin{figure}[ht!]
    \centering
    \begin{minipage}[b]{0.48\textwidth}
        \centering
        \includegraphics[width=\textwidth]{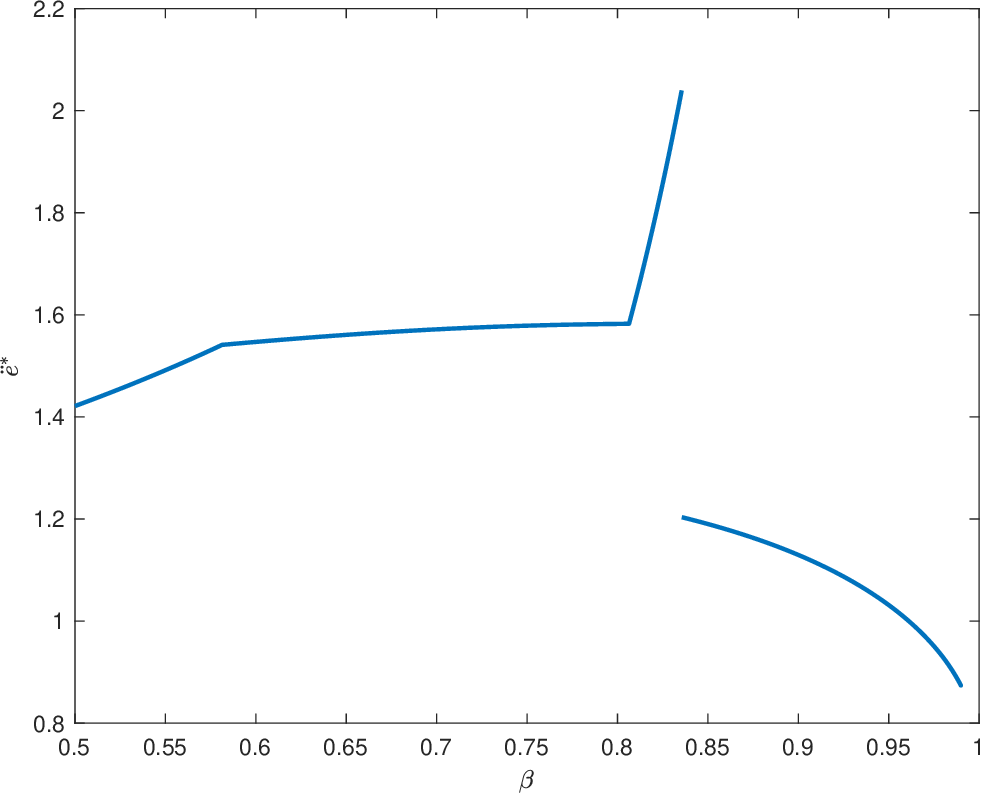}
        \caption{Optimal $\ddot{e}^*$ w.r.t. $\beta$}
        \label{fig:MHQ3e}
    \end{minipage}
    \hfill
    \begin{minipage}[b]{0.48\textwidth}
        \centering
        \includegraphics[width=\textwidth]{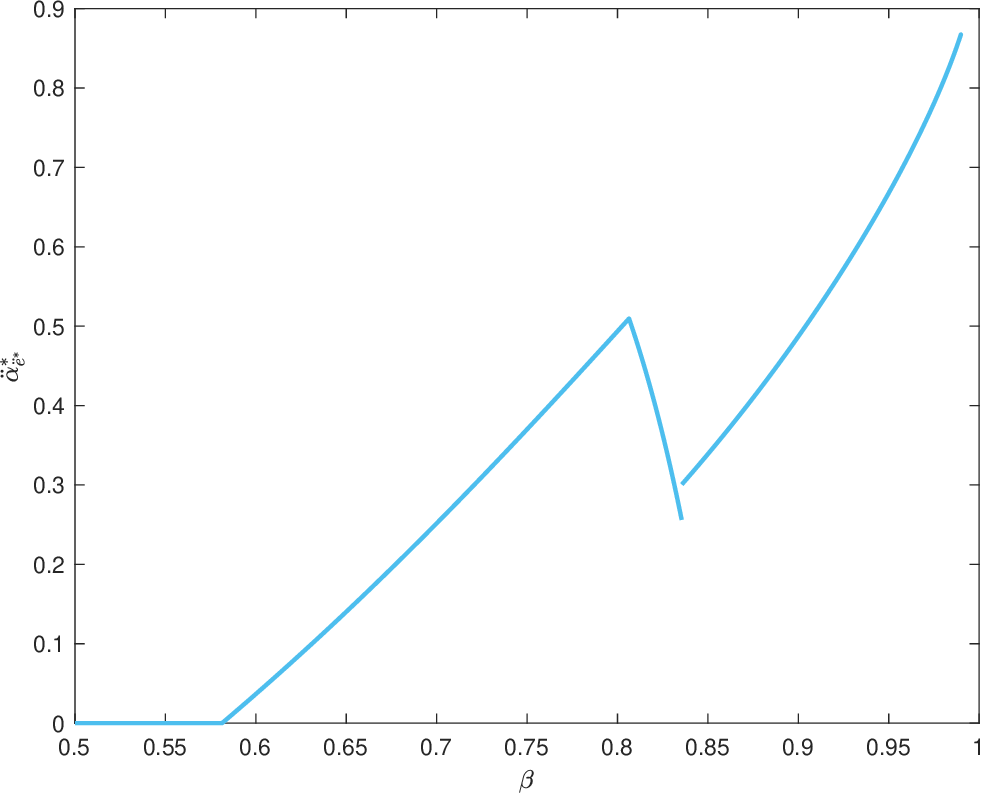}
        \caption{Optimal $\ddot{\alpha}_{\ddot{e}^*}^*$ w.r.t. $\beta$}
        \label{fig:MHQ3alpha}
    \end{minipage}
\end{figure}

Figure \ref{fig:MHQ3e} and \ref{fig:MHQ3alpha} illustrate the impact of the parameter $\beta$ on the optimal effort level $\ddot{e}^*$ and the optimal insurance coverage $\ddot{\alpha}_{\ddot{e}^*}^*$, respectively. 
From Figure \ref{fig:MHQ3e}, we can observe that as $\beta$ increases, the optimal effort level $\ddot{e}^*$ exhibits nonlinear changes: 
in the range $\beta \in [0.55, 0.83]$, $\ddot{e}^*$ steadily rises, indicating that individuals gradually increase their effort to manage risk; when $\beta > 0.83$, $\ddot{e}^*$ experiences a sharp jump to a higher level, followed by a declining branch. 
This pattern may reflect diminishing marginal returns of effort as $\beta$ increases, where individuals' willingness to exert effort diminishes at higher levels of risk aversion.
Figure \ref{fig:MHQ3alpha} illustrates that the optimal insurance demand $\ddot{\alpha}_{\ddot{e}^*}^*$ increases monotonically with $\beta$, indicating that higher $\beta$ values prompt individuals to rely more on insurance to mitigate risk exposure. 
Together with the previous figure, these results highlight the substitutability between effort and insurance. In the lower range of $\beta$, effort and insurance complement each other, working together to manage risk. However, in the higher range of $\beta$, insurance gradually becomes the primary tool for risk management, while the role of effort diminishes.

\section{Concluding remarks}\label{sec:conclusion}

Self-protection is commonly employed in insurance practice to help individuals mitigate risk by reducing the probability of claims. 
In this paper, we revisit the study of optimal insurance demand and self-protection efforts within the framework of DRMs. 
By utilizing convex premium principles, we derive optimal strategies for both insurance demand and prevention efforts when the insured adopts TVaR and strictly convex DRMs, respectively. 
Our findings underscore the complementary relationship between market insurance and self-protection. 
Additionally, we explore the impact of moral hazard on the proposed model. 
Our results show that moral hazard significantly reduces the insured's prevention efforts while altering the optimal insurance structure. 

Several promising directions for future research can be identified. 
First, it would be valuable to examine the interaction between market insurance and self-protection within broader frameworks, such as rank-dependent expected utility. 
Second, a deeper analysis of the objectives of both insured individuals and insurers could provide more insights into their interactions in the context of market insurance and self-protection.


\section*{Acknowledgements}
\noindent Yiying Zhang acknowledges the financial support from the GuangDong Basic and Applied Basic Research Foundation (No. 2023A1515011806), and Shenzhen Science and Technology Program (No. JCYJ20230807093312026). 

\section*{Disclosure statements}
\noindent No potential competing interests were reported by the authors.

\setlength{\bibsep}{0.0pt}
\bibliographystyle{mystyle}
\bibliography{prevention_convex}

\appendix

\section{Appendix}

In this appendix, we provide detailed proofs for all the results presented in the paper.

\subsection{Proof of Theorem \ref{thm:alphae}}
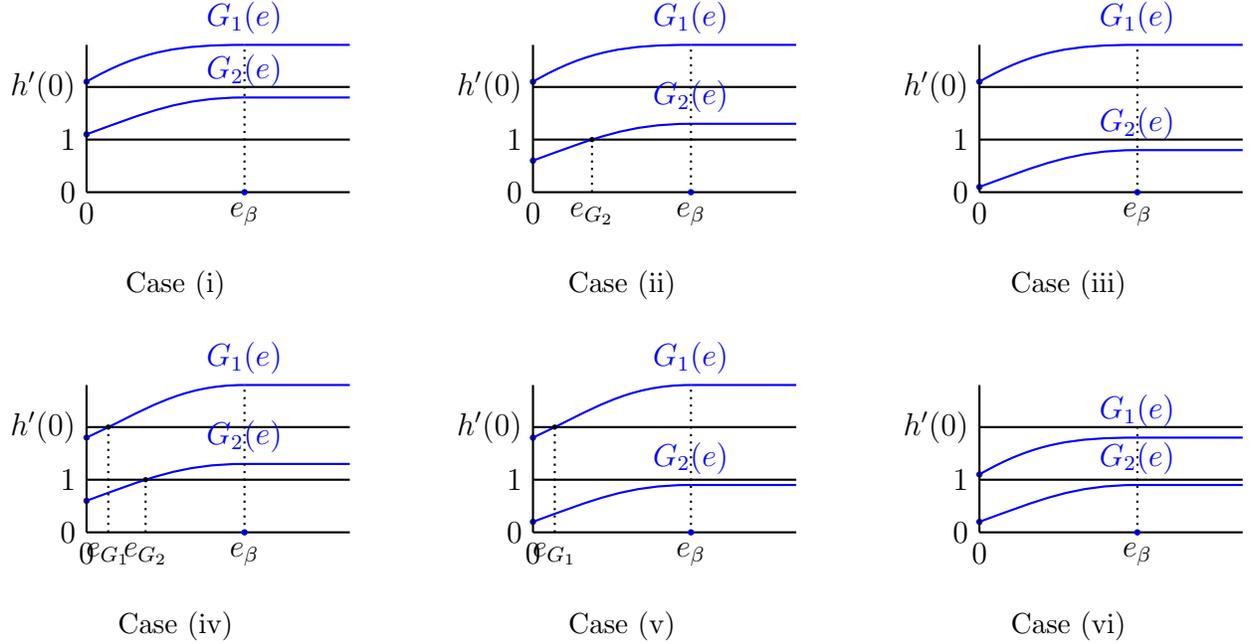
\begin{figure}[ht!]
\centering
\begin{tikzpicture}
  \matrix (m) [matrix of nodes, row sep=1cm, column sep=1cm]{
    \node (n11) {%
      \begin{tikzpicture}[scale=0.7]  
        \draw[thick] (0,0) -- (5,0);
        \draw[thick] (0,1) -- (5,1);
        \draw[thick] (0,2) -- (5,2);
        \node[left] at (0,2) {$h'(0)$};
        \node[left] at (0,1) {1};
        \node[left] at (0,0) {0};
        \node[below] at (0,0) {0};
        \node[below] at (3,0) {$e_\beta$};
        \draw[thick, blue] (0,2.1) to[out=30, in=180] (3,2.8) node[right, above] {$G_1(e)$};
        \draw[thick, blue] (0,1.1) to[out=20, in=180] (3,1.8) node[right, above] {$G_2(e)$};
        \draw[thick, blue] (3,1.8) to (5,1.8);
        \draw[thick, blue] (3,2.8) to (5,2.8);
        \filldraw[blue] (3,0) circle(.05)
        (0,1.1) circle(.05)
        (0,2.1) circle(.05);
        \draw[dotted, thick, black] (3,0) -- (3,2.8);
        \draw[thick, black] (0,0) -- (0,2.8);
      \end{tikzpicture}
    }; &
    \node (n12) {%
      \begin{tikzpicture}[scale=0.7]  
        \draw[thick] (0,0) -- (5,0);
        \draw[thick] (0,1) -- (5,1);
        \draw[thick] (0,2) -- (5,2);
        \node[left] at (0,2) {$h'(0)$};
        \node[left] at (0,1) {1};
        \node[left] at (0,0) {0};
        \node[below] at (0,0) {0};
        \node[below] at (3,0) {$e_\beta$};
        \draw[thick, blue] (0,2.1) to[out=30, in=180] (3,2.8) node[right, above] {$G_1(e)$};
        \draw[thick, blue] (0,0.6) to[out=20, in=180] (3,1.3) node[right, above] {$G_2(e)$};
        \path[name path=G2] (0,0.6) to[out=20, in=180] (3,1.3);
        \path[name path=line1] (0,1) to (5,1);
        \path[name intersections={of=G2 and line1, by=II}];
        \fill[black] (II) circle (.05);
        \draw[thick, blue] (3,2.8) to (5,2.8);
        \draw[thick, blue] (3,1.3) to (5,1.3);
        \filldraw[blue] (3,0) circle(.05)
        (0,0.6) circle(.05)
        (0,2.1) circle(.05);
        \draw[dotted, thick, black] (3,0) -- (3,2.8);
        \draw[thick, black] (0,0) -- (0,2.8);
        \draw[dotted, thick, black] (II|-0,0) -- (II);
        \node[below] at (II|-0,0) {$e_{G_2}$};
      \end{tikzpicture}
    }; &
    \node (n13) {%
      \begin{tikzpicture}[scale=0.7] 
        \draw[thick] (0,0) -- (5,0);
        \draw[thick] (0,1) -- (5,1);
        \draw[thick] (0,2) -- (5,2);
        \node[left] at (0,2) {$h'(0)$};
        \node[left] at (0,1) {1};
        \node[left] at (0,0) {0};
        \node[below] at (0,0) {0};
        \node[below] at (3,0) {$e_\beta$};
        \draw[thick, blue] (0,2.1) to[out=30, in=180] (3,2.8) node[right, above] {$G_1(e)$};
        \draw[thick, blue] (0,0.1) to[out=20, in=180] (3,0.8) node[right, above] {$G_2(e)$};
        \draw[thick, blue] (3,2.8) to (5,2.8);
        \draw[thick, blue] (3,0.8) to (5,0.8);
        \filldraw[blue] (3,0) circle(.05)
        (0,0.1) circle(.05)
        (0,2.1) circle(.05);
        \draw[dotted, thick, black] (3,0) -- (3,2.8);
        \draw[thick, black] (0,0) -- (0,2.8);
      \end{tikzpicture}
    }; \\
    \node (n21) {%
      \begin{tikzpicture}[scale=0.7]
        \draw[thick] (0,0) -- (5,0);
        \draw[thick] (0,1) -- (5,1);
        \draw[thick] (0,2) -- (5,2);
        \node[left] at (0,2) {$h'(0)$};
        \node[left] at (0,1) {1};
        \node[left] at (0,0) {0};
        \node[below] at (0,0) {0};
        \node[below] at (3,0) {$e_\beta$};
        \draw[thick, blue] (0,1.8) to[out=25, in=180] (3,2.8) node[right, above] {$G_1(e)$};
        \draw[thick, blue] (0,0.6) to[out=20, in=180] (3,1.3) node[right, above] {$G_2(e)$};
        \path[name path=G1] (0,1.8) to[out=25, in=180] (3,2.8);
        \path[name path=G2] (0,0.6) to[out=20, in=180] (3,1.3);
        \path[name path=line1] (0,1) to (5,1);
        \path[name path=line2] (0,2) to (5,2);
        \path[name intersections={of=G1 and line2, by=I}];
        \path[name intersections={of=G2 and line1, by=II}];
        \fill[black] (I) circle (.05);
        \fill[black] (II) circle (.05);
        \draw[thick, blue] (3,2.8) to (5,2.8);
        \draw[thick, blue] (3,1.3) to (5,1.3);
        \filldraw[blue] (3,0) circle(.05)
        (0,0.6) circle(.05)
        (0,1.8) circle(.05);
        \draw[dotted, thick, black] (3,0) -- (3,2.8);
        \draw[thick, black] (0,0) -- (0,2.8);
        \draw[dotted, thick, black] (I|-0,0) -- (I);
        \draw[dotted, thick, black] (II|-0,0) -- (II);
        \node[below] at (I|-0,0) {$e_{G_1}$};
        \node[below] at (II|-0,0) {$e_{G_2}$};
      \end{tikzpicture}
    }; &
    \node (n22) {%
      \begin{tikzpicture}[scale=0.7] 
        \draw[thick] (0,0) -- (5,0);
        \draw[thick] (0,1) -- (5,1);
        \draw[thick] (0,2) -- (5,2);
        \node[left] at (0,2) {$h'(0)$};
        \node[left] at (0,1) {1};
        \node[left] at (0,0) {0};
        \node[below] at (0,0) {0};
        \node[below] at (3,0) {$e_\beta$};
        \draw[thick, blue] (0,1.8) to[out=25, in=180] (3,2.8) node[right, above] {$G_1(e)$};
        \draw[thick, blue] (0,0.2) to[out=20, in=180] (3,0.9) node[right, above] {$G_2(e)$};
        \path[name path=G1] (0,1.8) to[out=25, in=180] (3,2.8);
        \path[name path=line2] (0,2) to (5,2);
        \path[name intersections={of=G1 and line2, by=I}];
        \fill[black] (I) circle (.05);
        \draw[thick, blue] (3,2.8) to (5,2.8);
        \draw[thick, blue] (3,0.9) to (5,0.9);
        \filldraw[blue] (3,0) circle(.05)
        (0,0.2) circle(.05)
        (0,1.8) circle(.05);
        \draw[dotted, thick, black] (3,0) -- (3,2.8);
        \draw[thick, black] (0,0) -- (0,2.8);
        \draw[dotted, thick, black] (I|-0,0) -- (I);
        \node[below] at (I|-0,0) {$e_{G_1}$};
      \end{tikzpicture}
    }; &
    \node (n23) {%
      \begin{tikzpicture}[scale=0.7]
        \draw[thick] (0,0) -- (5,0);
        \draw[thick] (0,1) -- (5,1);
        \draw[thick] (0,2) -- (5,2);
        \node[left] at (0,2) {$h'(0)$};
        \node[left] at (0,1) {1};
        \node[left] at (0,0) {0};
        \node[below] at (0,0) {0};
        \node[below] at (3,0) {$e_\beta$};
        \draw[thick, blue] (0,1.1) to[out=30, in=180] (3,1.8) node[right, above] {$G_1(e)$};
        \draw[thick, blue] (0,0.2) to[out=20, in=180] (3,0.9) node[right, above] {$G_2(e)$};
        \draw[thick, blue] (3,1.8) to (5,1.8);
        \draw[thick, blue] (3,0.9) to (5,0.9);
        \filldraw[blue] (3,0) circle(.05)
        (0,0.2) circle(.05)
        (0,1.1) circle(.05);
        \draw[dotted, thick, black] (3,0) -- (3,2);
        \draw[thick, black] (0,0) -- (0,2.8);
      \end{tikzpicture}
    }; \\
  };
  \node[below=5pt of n11] {\small Case (i)};
  \node[below=5pt of n12] {\small Case (ii)};
  \node[below=5pt of n13] {\small Case (iii)};
  \node[below=5pt of n21] {\small Case (iv)};
  \node[below=5pt of n22] {\small Case (v)};
  \node[below=5pt of n23] {\small Case (vi)};
\end{tikzpicture}
\caption{Schematic diagram of $G_1(e)$ and $G_2(e)$}
\label{fig:G1G2}
\end{figure}

It suffices to specify the sets of  $\mathcal{A}_1$ and $\mathcal{A}_2$ as given in (\ref{expr:A1A2}). 
Note that these two sets depend on the relationships between $G_1(e)$ and $h'(0)$ and between $G_2(e)$ and $1$. 
Figure \ref{fig:G1G2} presents all possible plots of $G_1(e)$ and $G_2(e)$ regarding to the locations of $h'(0)$ and 1. Next, we will discuss each case in detail to find out the exact solutions.

\underline{Case (i): $G_1(0)>h'(0)$ and $G_2(0)>1$}. As shown in Figure \ref{fig:G1G2}(i), we can immediately get $\mathcal{A}_1=\emptyset$ and  $\mathcal{A}_2=\mathbb{R}_+$. Hence $\alpha_e^*\equiv1$. 

\underline{Case (ii): $G_1(0)>h'(0)$ and $G_2(0)\leq1\leq G_2(e_\beta)$}.  
    As displayed in Figure \ref{fig:G1G2}(ii), we can get $\mathcal{A}_1=\emptyset$, $\mathcal{A}_2=[e_{G_2},+\infty)$, where $e_{G_2}$ is the solution of the equation $G_2(e)=1$, for $e\in [0,e_\beta]$. Hence, $$\alpha_e^*=\begin{cases}
        \alpha_{h,e}^*, & e\in[0,e_{G_2});\\
        1, & e\in [e_{G_2},+\infty),
    \end{cases}$$ where $\alpha_{h,e}^*$ satisfies equation (\ref{expr:alphahe}). 

\underline{Case (iii): $G_1(0)>h'(0)$ and $G_2(e_\beta)< 1$}. 
    As shown in Figure \ref{fig:G1G2}(iii), we can get $\mathcal{A}_1=\emptyset$ and $\mathcal{A}_2=\emptyset$. This means that the solution must be an inner point in $(0,1)$. Referring to equation (\ref{expr:alphahe}), we know $\alpha_{h,e}^*$ is the solution and well-defined on $[0,e_{\beta}]$. For $e\geq e_{\beta}$, it can be noted that equation (\ref{expr:alphahe}) has the following form
    \begin{equation*}
        \mathbb{E}[Y]=\int_{\beta(0)}^1\text{VaR}_s(Y)\dif s=(1-\beta)\mathbb{E}[Yh'(\alpha Y)],\quad\mbox{for $\alpha\in(0,1)$},
    \end{equation*}
    whose solution is denoted by $\alpha_{h,e_\beta}^*$. Besides, it is not hard to observe that $\lim_{e\to e_\beta^{-}}\alpha_{h,e}^*=\alpha_{h,e_\beta}^*$. Hence, the solution is given by 
    $$\alpha_e^*=\begin{cases}
        \alpha_{h,e}^*, & e\in[0,e_\beta);\\
        \alpha_{h,e_\beta}^*, & e\in[e_\beta,+\infty).
    \end{cases}$$

\underline{Case (iv): $G_1(0)\leq h'(0)\leq G_1(e_\beta)$ and $G_2(0)\leq 1\leq G_2(e_\beta)$}. 
    As illustrated in Figure \ref{fig:G1G2}(iv), we can get $\mathcal{A}_1=[0,e_{G_1}]$ and $\mathcal{A}_2=[e_{G_2},e_\beta)$, where $e_{G_1}$ is the solution of the equation $G_1(e)=h'(0)$, for $e\in [0,e_\beta)$. Hence, $$\alpha_e^*=\begin{cases}
        0, & e\in [0,e_{G_1}];\\
        \alpha_{h,e}^*, & e\in(e_{G_1},e_{G_2});\\
        1, & e\in [e_{G_2},+\infty).
    \end{cases}$$ 

\underline{Case (v): $G_1(0)\leq h'(0)\leq G_1(e_\beta)$ and $G_2(e_\beta)<1$}. 
    As shown in Figure \ref{fig:G1G2}(v), we can get $\mathcal{A}_1=[0,e_{G_1}]$ and $\mathcal{A}_2=\emptyset$. Hence, with a similar discussion in case (iii), one has
    $$\alpha_e^*=\begin{cases}
        0, & e\in [0,e_{G_1}];\\
        \alpha_{h,e}^*, & e\in(e_{G_1},e_\beta);\\
        \alpha_{h,e_\beta}, &e\in[e_\beta,+\infty).
    \end{cases}$$

\underline{Case (vi): $G_1(e_\beta)< h'(0)$ and $G_2(e_\beta)<1$}. As displayed in Figure \ref{fig:G1G2}(vi), we can get $\mathcal{A}_1=\mathbb{R}_+$ and $\mathcal{A}_2=\emptyset$. Hence, $\alpha_e^*=0$, for $e\in\mathbb{R}_+$.

Next, we show that $\alpha_{h,e}^*$ is continuous and non-decreasing in $e$ whenever it appears for the above-considered cases. Note that $\alpha_{h,e}^*$ is the solution of the equation
$$\int_{\beta(e)}^1\text{VaR}_s(Y)\dif s=(1-\beta)\mathbb{E}[Y\cdot h'(\alpha_{h,e}^*Y)],$$
for $e\in[0,e_\beta]$. Since $\beta(e)$ is a continuous and strictly decreasing function of $e$, then for a non-negative continuous random variable $Y$ and strictly convex $h\in\mathcal{H}$, we can infer that the left-hand side of the above equation is continuous and increasing on $e\in[0,e_\beta]$. Hence, $\alpha_{h,e}^*$ must be continuous and increasing in $e\in[0,e_\beta]$.

On the other hand, for $e\in[0,e_\beta]$, if either the solution of $G_1(e)=h'(0)$ exists (that is $e_{G_1}$), or the solution of $G_2(e)=1$ exits (that is $e_{G_2}$), or both hold, we must have  have $\lim_{e\rightarrow e_{G_1}^+} \alpha_{h,e}^*=\alpha_{h,e_{G_1}}^*$ and $\lim_{e\rightarrow e_{G_2}^-} \alpha_{h,e}^*=\alpha_{h,e_{G_2}}^*$ by the continuity of $\alpha_{h,e}^*$. 

We next show that $\alpha_{h,e_{G_1}}^*=0$ and $\alpha_{h,e_{G_2}}^*=1$. For $\alpha_{h,e_{G_1}}^*$, we know that
\begin{equation*}
G_1(e_{G_1})=\frac{\int_{\beta(e_{G_1})}^1\text{VaR}_s(Y)\dif s}{(1-\beta)\mathbb{E}[Y]}=h'(0),
\end{equation*}
which means that 
$$(1-\beta)\mathbb{E}[h'(0)Y]=\int_{\beta(e_{G_1})}^1\text{VaR}_s(Y)\dif s=(1-\beta)\mathbb{E}[Y\cdot h'(\alpha_{h,e_{G_1}}^* Y)].$$ 
By the properties of $h'$ and $Y$, we can get that $\lim_{e\rightarrow e_{G_1}^+} \alpha_{h,e}^*=\alpha_{h,e_{G_1}}^*=0$. 
Similarly, $\lim_{e\rightarrow e_{G_2}^-} \alpha_{h,e}^*=\alpha_{h,e_{G_2}}^*=1.$\hfill $\blacksquare$

\subsection{Proof of Theorem \ref{thm:e}}
According to the classifications in Theorem \ref{thm:alphae}, the solution of the optimal $e^*$ can be conducted as follows:

\underline{Case (i): $G_1(0)>h'(0)$ and $G_2(0)>1$}.  As per Theorem \ref{thm:alphae}, the object function $K_1(e)$ under this case is given by
        \begin{equation}\label{expr:K1}
            K_1(e)=p(e)\mathbb{E}[h(Y)]+c(e).
        \end{equation}
Since $p(e)$ and $c(e)$ are both convex, $K_1(e)$ is a convex function, hence its first derivative $K_1'(e)$ is an increasing function. On the other hand, based on Assumptions \ref{protection:e}-\ref{cost:e}, we have $K_1'(0)<0$ and $K_1'(+\infty)=+\infty$. Therefore, there must exist a unique interior point $\hat{e}$ in the interval $e\in[0, +\infty)$ such that $K_1'(\hat{e}) = 0$. Hence, $K_1(e)$ is firstly decreasing on $[0, \hat{e}]$ and then increasing on $[\hat{e}, +\infty)$. Therefore, $K_1(e)$ attains its minimum value at $\hat{e}$. This proof is finished.

\underline{Case (ii): $G_1(0)>h'(0)$ and $G_2(0)\leq 1\leq G_2(e_\beta)$}. In light of Theorem \ref{thm:alphae}, the objective function $K(e)$ has the following form:
        \begin{equation}\label{expr:K2}
            K_2(e)=\begin{cases}
                p(e)\mathbb{E}[h(\alpha_{h,e}^*Y)]+\frac{(1-\alpha_{h,e}^*)p(e)}{1-\beta}\int_{\beta(e)}^1 \text{VaR}_s(Y)\dif s+c(e), & e\in[0,e_{G_2});\\
                p(e)\mathbb{E}[h(Y)]+c(e), & e\in[e_{G_2},+\infty).         
                \end{cases}
        \end{equation}
By noting that $\alpha_{h,e_{G_2}}^*=1$ (which has been proven in Theorem \ref{thm:alphae}), $K_2(e)$ is a continuous function. For $e\in[e_{G_2},+\infty)$, $K_2(e)$ is strictly convex as argued in Case (i). Therefore, there must exist a point $\hat{e}$, such that $\hat{e}={\arg\min}_{e\in[e_{G_2},+\infty)}K_2(e)$. 

However, for $e\in[0,e_{G_2})$, since the monotonicity and convexity properties of $K_2(e)$ in this interval are unclear, we consider the following two cases: (a) $K_2(e)$ does not attain its minimum value within this interval, i.e., $e_{G_2}={\arg\min}_{e\in[0, e_{G_2}]} K_2(e)$, as $K_2(e)$ is continuous at $e_{G_2}$; (b) there exists an interior point $\tilde{e} \in [0, e_{G_2})$ such that $\tilde{e}={\arg\min}_{e\in[0, e_{G_2})} K_2(e)$.
\begin{enumerate}
            \item[(a)] if $e_{G_2}={\arg\min}_{e\in[0, e_{G_2}]} K_2(e)$, we just need to find out $\hat{e}$, and the global minimum is attained at $\hat{e}$;
            \item[(b)] if $\tilde{e}={\arg\min}_{e\in[0, e_{G_2})} K_2(e)$, then we must compare the values of $K_2(\tilde{e})$ and $K_2(\hat{e})$:
            \begin{itemize}
                \item if $K_2(\tilde{e}) \leq K_2(\hat{e})$, then the global minimum is attained at $\tilde{e}$;
                \item if $K_2(\tilde{e}) > K_2(\hat{e})$, then the global minimum is attained at $\hat{e}$.
            \end{itemize}
\end{enumerate}
Hence, the proof is completed.

\underline{Case (iii): $G_1(0)>h'(0)$ and $G_2(e_\beta)<1$}. It is not hard to check that
        \begin{equation}\label{expr:K3}
            K_3(e)=\begin{cases}
                p(e)\mathbb{E}[h(\alpha_{h,e}^*Y)]+\frac{(1-\alpha_{h,e}^*)p(e)}{1-\beta}\int_{\beta(e)}^1 \text{VaR}_s(Y)\dif s+c(e), & e\in[0,e_\beta);\\
                p(e)\mathbb{E}[h(\alpha_{h,e_\beta}^*Y)]+\frac{(1-\alpha_{h,e_\beta}^*)p(e)}{1-\beta}\mathbb{E}[Y]+c(e), & e\in[e_\beta,+\infty).         
                \end{cases}
        \end{equation}
Since $\beta(e_\beta)=0$, $K_3(e)$ is a continuous function. For $e\in[e_\beta,+\infty)$, $\alpha_{h,e_\beta}^*$ is a constant independent of $e$, which takes value in $(0,1)$, $K_3(e)$ must be strictly convex on this interval. Therefore, there must exists a point $\text{\r e}$ such that $\text{\r e}={\arg\min}_{e\in[e_\beta,+\infty)}K_3(e)$. On the other hand, for $e\in[0,e_\beta)$, the monotonicity and convexity properties of $K_3(e)$ in this interval are unknown, we consider the following two cases: (a) $K_3(e)$ does not attain its minimum value within this interval, i.e., $e_\beta={\arg\min}_{e\in[0, e_\beta]} K_3(e)$, as $K_3(e)$ is continuous at $e_\beta$; (b) there exists an interior point $\tilde{e} \in [0, e_\beta)$ such that $\tilde{e}={\arg\min}_{e\in[0, e_\beta)} K_3(e)$.
        \begin{enumerate}
            \item[(a)] if $e_\beta={\arg\min}_{e\in[0, e_\beta]} K_3(e)$, we just need to find $\text{\r e}$, and the global minimum is attained at $\text{\r e}$;
            \item[(b)] if $\tilde{e}={\arg\min}_{e\in[0, e_\beta)} K_3(e)$, then we must compare the values of $K_3(\tilde{e})$ and $K_3(\text{\r e})$:
            \begin{itemize}
                \item if $K_3(\tilde{e}) \leq K_3(\text{\r e})$, then the global minimum is attained at $\tilde{e}$;
                \item if $K_3(\tilde{e}) > K_3(\text{\r e})$, then the global minimum is attained at $\text{\r e}$.
            \end{itemize}
        \end{enumerate}
Hence, the proof for this case is finished.

\underline{Case (iv): $G_1(0)\leq h'(0)\leq G_1(e_\beta)$ and $G_2(0)\leq1\leq G_2(e_\beta)$}. 
        From Theorem \ref{thm:alphae}, the objective function in (\ref{expr:Ke}) can be simplified as
        \begin{equation}\label{expr:K4}
            K_4(e)=\begin{cases}
                \frac{p(e)}{1-\beta}\int_{\beta(e)}^1\text{VaR}_s(Y) \dif s+c(e), &e\in[0,e_{G_1}];\\
                p(e)\mathbb{E}[h(\alpha_{h,e}^*Y)]+\frac{(1-\alpha_{h,e}^*)p(e)}{1-\beta}\int_{\beta(e)}^1 \text{VaR}_s(Y)\dif s+c(e), & e\in(e_{G_1},e_{G_2});\\
                p(e)\mathbb{E}[h(Y)]+c(e), & e\in[e_{G_2},+\infty).         
                \end{cases}
        \end{equation}
According to the proof of Theorem \ref{thm:alphae}, we know that $\alpha_{h,e_{G_1}}^*=0$ and $\alpha_{h,e_{G_2}}^*=1$, which implies that $K_4(e)$ is continuous on $\mathbb{R}_+$. For the intervals $[0,e_{G_1}]$ and $[e_{G_2},+\infty)$, $K_4(e)$ is strictly convex on both intervals respectively in accordance with Assumptions \ref{protection:e}, \ref{cost:e} and \ref{ass:convexTVaR}. 
As a result, we let $\Dot{e}$ denote the point where $K_4(e)$ attains its minimum in the interval $[0, e_{G_1}]$, and let $\hat{e}$ denote the point where $K_4(e)$ attains its minimum in the interval $[e_{G_2}, +\infty)$. 

However, for the interval $e\in(e_{G_1},e_{G_2})$, the monotonicity or convexity properties of $K_4(e)$ are hard to infer. Next, we will analyze the minimum value by assuming that whether there is an interior minimum point in $(e_{G_1},e_{G_2)}$. Based on Assumption \ref{ass:uniqueTVaR}, the discussion is divided into the following three cases:
        \begin{enumerate}
            \item[(a)] If $e_{G_1}={\arg\min}_{e\in[e_{G_1},e_{G_2}]}K_4(e)$, then on the interval $[0,e_{G_2})$, the minimum is attained at $\Dot{e}$, and we need to compare the values of $K_4(\Dot{e})$ and $K_4(\hat{e})$:
            \begin{itemize}
                \item if $K_4(\Dot{e})\leq K_4(\hat{e})$, then the global minimum is attained at $\Dot{e}$;
                \item if $K_4(\Dot{e})> K_4(\hat{e})$, then the global minimum is attained at $\hat{e}$.
            \end{itemize}
            \item[(b)] If $e_{G_2}={\arg\min}_{e\in[e_{G_1},e_{G_2}]}K_4(e)$, then on the interval $(e_{G_1},+\infty)$, the minimum is attained at $\hat{e}$, and we need to compare the values of $K_4(\Dot{e})$ and $K_4(\hat{e})$:
            \begin{itemize}
                \item if $K_4(\Dot{e})\leq K_4(\hat{e})$, then the global minimum is attained at $\Dot{e}$;
                \item if $K_4(\Dot{e})> K_4(\hat{e})$, then the global minimum is attained at $\hat{e}$.
            \end{itemize}
            \item[(c)] If there exists an interior point $\tilde{e}$ such that  $\tilde{e}={\arg\min}_{e\in(e_{G_1},e_{G_2})}K_4(e)$, then we need to compare $K_4(\Dot{e})$, $K_4(\tilde{e})$ and $K_4(\hat{e})$:
            \begin{itemize}
                \item if $K_4(\Dot{e})\leq\min\{K_4(\tilde{e}), K_4(\hat{e})\}$, then the global minimum is attained at $\Dot{e}$;
                \item if $K_4(\tilde{e})\leq\min\{K_4(\Dot{e}), K_4(\hat{e})\}$, then the global minimum is attained at $\tilde{e}$;
                \item if $K_4(\hat{e})\leq\min\{K_4(\Dot{e}), K_4(\tilde{e})\}$, then the global minimum is attained at $\hat{e}$.
            \end{itemize}
        \end{enumerate}
To sum up, the proof for this case is finished.

\underline{Case (v): $G_1(0)\leq h'(0)\leq G_1(e_\beta)$ and $G_2(e_\beta)<1$}. For this case, the objective function $K(e)$ has the following expression:
        \begin{equation}\label{expr:K5}
            K_5(e)=\begin{cases}
                \frac{p(e)}{1-\beta}\int_{\beta(e)}^1\text{VaR}_s(Y) \dif s+c(e), &e\in[0,e_{G_1}];\\
                p(e)\mathbb{E}[h(\alpha_{h,e}^*Y)]+\frac{(1-\alpha_{h,e}^*)p(e)}{1-\beta}\int_{\beta(e)}^1 \text{VaR}_s(Y)\dif s+c(e), & e\in(e_{G_1},e_\beta);\\
                p(e)\mathbb{E}[h(\alpha_{h,e_\beta}^*Y)]+\frac{(1-\alpha_{h,e_\beta}^*)p(e)}{1-\beta}\mathbb{E}[Y]+c(e), & e\in[e_\beta,+\infty).   
                \end{cases}
        \end{equation}
As shown earlier, we have $\alpha_{h,e_{G_1}}^*=0$ and $\beta(e_\beta)=0$, which means that $K_5(e)$ is continuous on $\mathbb{R}_+$. For the intervals $[0,e_{G_1}]$ and $[e_{G_2},+\infty)$, it is easy to see that $K_5(e)$ is strictly convex on both intervals respectively due to Assumptions \ref{protection:e}, \ref{cost:e} and \ref{ass:convexTVaR} and the fact that $\alpha_{h,e_\beta}^*$ is independent of $e$. For ease of discussions, we let $\Dot{e}$ denote the point where $K_5(e)$ attains its minimum in the interval $[0, e_{G_1}]$, and let $\text{\r e}$ denote the point where $K_5(e)$ attains its minimum in the interval $[e_\beta, +\infty)$. 

Now, we consider the objection function on $e\in(e_{G_1},e_\beta)$. Based on Assumption \ref{ass:uniqueTVaR}, the following subcases are considered:
        \begin{enumerate}
            \item[(a)] If $e_{G_1}={\arg\min}_{e\in[e_{G_1},e_\beta]}K_5(e)$, then on the interval $[0,e_\beta)$, the minimum is attained at $\Dot{e}$, we need to compare the values of $K_5(\Dot{e})$ and $K_5(\text{\r e})$ as follows:
            \begin{itemize}
                \item if $K_5(\Dot{e})\leq K_5(\text{\r e})$, then the global minimum is attained at $\Dot{e}$;
                \item if $K_5(\Dot{e})> K_5(\text{\r e})$, then the global minimum is attained at $\text{\r e}$.
            \end{itemize}
            \item[(b)] If $e_\beta={\arg\min}_{e\in[e_{G_1},e_\beta]}K_5(e)$, then in $(e_{G_1},+\infty)$, the minimum is attained at $\text{\r e}$, we just need to compare the values of $K_5(\Dot{e})$ and $K_5(\text{\r e})$:
            \begin{itemize}
                \item if $K_5(\Dot{e})\leq K_5(\text{\r e})$, then the global minimum is attained at $\Dot{e}$;
                \item if $K_5(\Dot{e})> K_5(\text{\r e})$, then the global minimum is attained at $\text{\r e}$.
            \end{itemize}
            \item[(c)] If there exists an interior point $\tilde{e}$ such that   $\tilde{e}={\arg\min}_{e\in(e_{G_1},e_\beta)}K_5(e)$, then we need to compare $K_5(\Dot{e})$, $K_5(\tilde{e})$ and $K_5(\text{\r e})$:
            \begin{itemize}
                \item if $K_5(\Dot{e})\leq\min\{K_5(\tilde{e}), K_5(\text{\r e})\}$, then the global minimum is attained at $\Dot{e}$;
                \item if $K_5(\tilde{e})\leq\min\{K_5(\Dot{e}), K_5(\text{\r e})\}$, then the global minimum is attained at $\tilde{e}$;
                \item if $K_5(\text{\r e})\leq\min\{K_5(\Dot{e}), K_5(\tilde{e})\}$, then the global minimum is attained at $\text{\r e}$.
            \end{itemize}
        \end{enumerate}
Therefore, the proof is finished for this case.

\underline{Case (vi): $G_1(e_\beta)<h'(0)$ and $G_2(e_\beta)<1$}. In this case, the objective function $K(e)$ has the following expression:
        \begin{equation}\label{expr:K6}
            K_6(e)=\begin{cases}
                \frac{p(e)}{1-\beta}\int_{\beta(e)}^1\text{VaR}_s(Y) \dif s+c(e), &e\in[0,e_\beta);\\
                \frac{p(e)}{1-\beta}\mathbb{E}[Y]+c(e), & e\in[e_\beta,+\infty).   
                \end{cases}
        \end{equation}
Since $\beta(e_{\beta})=0$, we know that $K_6(e)$ is continuous on $\mathbb{R}_+$. On both of the intervals $[0,e_\beta)$ and $[e_\beta,+\infty)$, $K_6(e)$ is strictly convex due to Assumptions \ref{protection:e}-\ref{ass:convexTVaR}. Let $\Dot{e}$ denote the point where $K_6(e)$ attains its minimum in the interval $[0, e_\beta)$, and let $\hat{e}$ denote the point where $K_6(e)$ attains its minimum in the interval $[e_\beta, +\infty)$. The optimal solution can be achieved by comparing $K_6(\Dot{e})$ and $K_6(\hat{e})$: if $K_6(\Dot{e})\leq K_6(\hat{e})$, then $e^*=\Dot{e}$; otherwise, $e^*=\hat{e}$. 
\hfill $\blacksquare$

\subsection{Proof of Theorem \ref{thm:alphae2}}
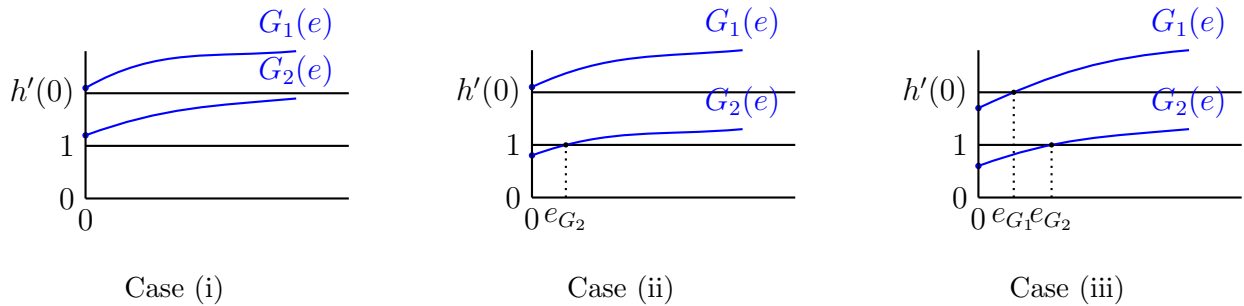
\begin{figure}[ht!]
\centering
\begin{tikzpicture}
  \matrix (m) [matrix of nodes, row sep=1cm, column sep=1cm]{
    \node (n11) {%
      \begin{tikzpicture}[scale=0.7]
        \draw[thick] (0,0) -- (5,0);
        \draw[thick] (0,1) -- (5,1);
        \draw[thick] (0,2) -- (5,2);
        \node[left] at (0,2) {$h'(0)$};
        \node[left] at (0,1) {1};
        \node[left] at (0,0) {0};
        \node[below] at (0,0) {0};
        \draw[thick, blue] (0,2.1) to[out=30, in=185] (4,2.8) node[right, above] {$G_1(e)$};
        \draw[thick, blue] (0,1.2) to[out=20, in=185] (4,1.9) node[right, above] {$G_2(e)$};
        \filldraw[blue] (0,1.2) circle(.05)
        (0,2.1) circle(.05);
        \draw[thick, black] (0,0) -- (0,2.8);
      \end{tikzpicture}
    }; &
    \node (n12) {%
      \begin{tikzpicture}[scale=0.7] 
        \draw[thick] (0,0) -- (5,0);
        \draw[thick] (0,1) -- (5,1);
        \draw[thick] (0,2) -- (5,2);
        \node[left] at (0,2) {$h'(0)$};
        \node[left] at (0,1) {1};
        \node[left] at (0,0) {0};
        \node[below] at (0,0) {0};
        \draw[thick, blue] (0,2.1) to[out=25, in=185] (4,2.8) node[right, above] {$G_1(e)$};
        \draw[thick, blue] (0,0.8) to[out=20, in=185] (4,1.3) node[right, above] {$G_2(e)$};
        \path[name path=G2] (0,0.8) to[out=20, in=185] (4,1.3);
        \path[name path=line2] (0,1) to (5,1);
        \path[name intersections={of=G2 and line2, by=I}];
        \fill[black] (I) circle (.05);
        \filldraw[blue] (0,0.8) circle(.05)
        (0,2.1) circle(.05);
        \draw[thick, black] (0,0) -- (0,2.8);
        \draw[dotted, thick, black] (I|-0,0) -- (I);
        \node[below] at (I|-0,0) {$e_{G_2}$};
      \end{tikzpicture}
    }; &
    \node (n13) {%
    \begin{tikzpicture}[scale=0.7]
        \draw[thick] (0,0) -- (5,0);
        \draw[thick] (0,1) -- (5,1);
        \draw[thick] (0,2) -- (5,2);
        \node[left] at (0,2) {$h'(0)$};
        \node[left] at (0,1) {1};
        \node[left] at (0,0) {0};
        \node[below] at (0,0) {0};
        \draw[thick, blue] (0,1.7) to[out=25, in=185] (4,2.8) node[right, above] {$G_1(e)$};
        \draw[thick, blue] (0,0.6) to[out=20, in=185] (4,1.3) node[right, above] {$G_2(e)$};
        \path[name path=G1] (0,1.7) to[out=25, in=185] (4,2.8);
        \path[name path=G2] (0,0.6) to[out=20, in=185] (4,1.3);
        \path[name path=line1] (0,1) to (5,1);
        \path[name path=line2] (0,2) to (5,2);
        \path[name intersections={of=G1 and line2, by=I}];
        \path[name intersections={of=G2 and line1, by=II}];
        \fill[black] (I) circle (.05);
        \fill[black] (II) circle (.05);
        \filldraw[blue] (0,0.6) circle(.05)
        (0,1.7) circle(.05);
        \draw[thick, black] (0,0) -- (0,2.8);
        \draw[dotted, thick, black] (I|-0,0) -- (I);
        \draw[dotted, thick, black] (II|-0,0) -- (II);
        \node[below] at (I|-0,0) {$e_{G_1}$};
        \node[below] at (II|-0,0) {$e_{G_2}$};
      \end{tikzpicture}
    }; \\
  };
  \node[below=5pt of n11] {\small Case (i)};
  \node[below=5pt of n12] {\small Case (ii)};
  \node[below=5pt of n13] {\small Case (iii)};
\end{tikzpicture}
\caption{Schematic diagram of $G_1(e)$ and $G_2(e)$}
\label{fig:G1G2-2}
\end{figure}

The proof is very similar to Theorem \ref{thm:alphae}, and here we provide for completeness. We need to specify the sets of  $\mathcal{A}_1$ and $\mathcal{A}_2$ given in (\ref{expr:A1A2}) by determining  the relationships between $G_1(e)$ and $h'(0)$ and between $G_2(e)$ and $1$. Figure \ref{fig:G1G2-2} presents all possible plots of $G_1(e)$ and $G_2(e)$ regarding to the locations of $h'(0)$ and 1. Under Assumptions \ref{protection:e} and \ref{ass:psi}, we know that $\psi(e)$ is non-decreasing in $e\in\mathbb{R}_+$. Without loss of generality, we assume that  $\lim_{e\to+\infty}\psi(e)=+\infty$. For the case when $\lim_{e\to+\infty}\psi(e)$ is finite, the discussions are very similar with those ones of Theorem \ref{thm:e} and thus omitted here for brevity.

\underline{Case (i): $G_1(0)>h'(0)$ and $G_2(0)>1$}. As plotted in Figure \ref{fig:G1G2-2}(i), one has  $\mathcal{A}_1=\emptyset$ and  $\mathcal{A}_2=\mathbb{R}_+$. Hence $\alpha_e^*\equiv1$.

\underline{Case (ii): $G_1(0)>h'(0)$ and $G_2(0)\leq1$}.  
    As displayed in Figure \ref{fig:G1G2-2}(ii), we can get $\mathcal{A}_1=\emptyset$, $\mathcal{A}_2=[e_{G_2},+\infty)$, where $e_{G_2}$ is the solution of the equation $G_2(e)=1$, for $e\in \mathbb{R}_+$. Hence, $$\alpha_e^*=\begin{cases}
        \alpha_{h,e}^*, & e\in[0,e_{G_2});\\
        1, & e\in [e_{G_2},+\infty),
    \end{cases}$$ where $\alpha_{h,e}^*$ satisfies equation (\ref{expr:alphahe2}).

\underline{Case (iii): $G_1(0)\leq h'(0)$ and $G_2(0)\leq 1$}. 
    As illustrated in Figure \ref{fig:G1G2-2}(iii), we can get $\mathcal{A}_1=[0,e_{G_1}]$ and $\mathcal{A}_2=[e_{G_2},e_\beta)$, where $e_{G_1}$ is the solution of the equation $G_1(e)=h'(0)$, for $e\in\mathbb{R}_+$. Hence, $$\alpha_e^*=\begin{cases}
        0, & e\in [0,e_{G_1}];\\
        \alpha_{h,e}^*, & e\in(e_{G_1},e_{G_2});\\
        1, & e\in [e_{G_2},+\infty).
    \end{cases}$$ 

Next, we show that $\alpha_{h,e}^*$ is continuous and non-decreasing in $e$ for cases (ii) and (iii). Note that $\alpha_{h,e}^*$ is the solution of the equation
$$\mathbb{E}[Y\cdot h'(\alpha_{h,e}^*Y)]=\frac{\int_0^{p(e)}\text{VaR}_{1-\frac{u}{p(e)}(Y)}\dif g(u)}{p(e)}=\psi(e),$$
for $e\in\mathbb{R}_+$. By Assumption \ref{ass:psi}, we have already known the right-hand side of the above equation is continuous and increasing on $e\in\mathbb{R}_+$. Then for a non-negative continuous random variable $Y$ and strictly convex $h\in\mathcal{H}$, $\alpha_{h,e}^*$ must be continuous and increasing in $e\in\mathbb{R}_+$.

On the other hand, for $e\in\mathbb{R}_+$, if either the solution of $G_1(e)=h'(0)$ exists (that is $e_{G_1}$), or the solution of $G_2(e)=1$ exits (that is $e_{G_2}$), or both hold, we must have  have $\lim_{e\rightarrow e_{G_1}^+} \alpha_{h,e}^*=\alpha_{h,e_{G_1}}^*$ and $\lim_{e\rightarrow e_{G_2}^-} \alpha_{h,e}^*=\alpha_{h,e_{G_2}}^*$ by the continuity of $\alpha_{h,e}^*$. 

We next show that $\alpha_{h,e_{G_1}}^*=0$ and $\alpha_{h,e_{G_2}}^*=1$. For $\alpha_{h,e_{G_1}}^*$, we know that
\begin{equation*}
G_1(e_{G_1})=\frac{\psi(e_{G_1})}{\mathbb{E}[Y]}=h'(0),
\end{equation*}
which means that 
$$\mathbb{E}[h'(0)Y]=\psi(e_{G_1})=\mathbb{E}[Y\cdot h'(\alpha_{h,e_{G_1}}^* Y)].$$ 
By the properties of $h'$ and $Y$, we can get that $\lim_{e\rightarrow e_{G_1}^+} \alpha_{h,e}^*=\alpha_{h,e_{G_1}}^*=0$. 
Similarly, $\lim_{e\rightarrow e_{G_2}^-} \alpha_{h,e}^*=\alpha_{h,e_{G_2}}^*=1.$ Hence, the proof is finished.
\hfill $\blacksquare$

\subsection{Proof of Theorem \ref{thm:e2}}
According to the classifications in Theorem \ref{thm:alphae2}, the solution of the optimal $e^*$ can be conducted as follows:

\underline{Case (i): $G_1(0)>h'(0)$ and $G_2(0)>1$}.  The proof is similar to Theorem \ref{thm:e}(i).


\underline{Case (ii): $G_1(0)>h'(0)$ and $G_2(0)\leq 1$}. In light of Theorem \ref{thm:alphae2}, the objective function $K(e)$ has the following form:
        \begin{equation}\label{expr:K2-2}
            K_2(e)=\begin{cases}
                p(e)\mathbb{E}[h(\alpha_{h,e}^*Y)]+(1-\alpha_{h,e}^*)\int_0^{p(e)}\text{VaR}_{1-\frac{u}{p(e)}}(Y)\dif g(u)+c(e), & e\in[0,e_{G_2});\\
                p(e)\mathbb{E}[h(Y)]+c(e), & e\in[e_{G_2},+\infty).         
                \end{cases}
        \end{equation}
By noting that $\alpha_{h,e_{G_2}}^*=1$ (which has been proven in Theorem \ref{thm:alphae2}), $K_2(e)$ is a continuous function. For $e\in[e_{G_2},+\infty)$, $K_2(e)$ is strictly convex as argued in Theorem \ref{thm:e}(i). Therefore, there must exist a point $\hat{e}$, such that $\hat{e}={\arg\min}_{e\in[e_{G_2},+\infty)}K_2(e)$. 

For $e\in[0,e_{G_2})$, we consider the following two cases: (a) $K_2(e)$ does not attain its minimum value within this interval, i.e., $e_{G_2}={\arg\min}_{e\in[0, e_{G_2}]} K_2(e)$, as $K_2(e)$ is continuous at $e_{G_2}$; (b) there exists an interior point $\tilde{e} \in [0, e_{G_2})$ such that $\tilde{e}={\arg\min}_{e\in[0, e_{G_2})} K_2(e)$.
\begin{enumerate}
            \item[(a)] if $e_{G_2}={\arg\min}_{e\in[0, e_{G_2}]} K_2(e)$, the global minimum is attained at $\hat{e}$;
            \item[(b)] if $\tilde{e}={\arg\min}_{e\in[0, e_{G_2})} K_2(e)$, then we must compare the values of $K_2(\tilde{e})$ and $K_2(\hat{e})$:
            \begin{itemize}
                \item if $K_2(\tilde{e}) \leq K_2(\hat{e})$, then the global minimum is attained at $\tilde{e}$;
                \item if $K_2(\tilde{e}) > K_2(\hat{e})$, then the global minimum is attained at $\hat{e}$.
            \end{itemize}
\end{enumerate}
Hence, the proof is completed.

\underline{Case (iii): $G_1(0)\leq h'(0)$ and $G_2(0)\leq1$}. 
        From Theorem \ref{thm:alphae2}, the objective function in (\ref{expr:Ke}) can be simplified as
        \begin{equation}\label{expr:K3-2}
            K_3(e)=\begin{cases}
                \int_0^{p(e)}\text{VaR}_{1-\frac{u}{p(e)}}(Y)\dif g(u)+c(e), &e\in[0,e_{G_1}];\\
                p(e)\mathbb{E}[h(\alpha_{h,e}^*Y)]+(1-\alpha_{h,e}^*)\int_0^{p(e)}\text{VaR}_{1-\frac{u}{p(e)}}(Y)\dif g(u)+c(e), & e\in(e_{G_1},e_{G_2});\\
                p(e)\mathbb{E}[h(Y)]+c(e), & e\in[e_{G_2},+\infty).         
                \end{cases}
        \end{equation}
According to the proof of Theorem \ref{thm:alphae2}, we know that $\alpha_{h,e_{G_1}}^*=0$ and $\alpha_{h,e_{G_2}}^*=1$, which implies that $K_3(e)$ is continuous on $\mathbb{R}_+$. For the intervals $[0,e_{G_1}]$ and $[e_{G_2},+\infty)$, $K_4(e)$ is strictly convex on both intervals respectively in accordance with Assumptions \ref{protection:e}, \ref{cost:e} and \ref{ass:convexTVaR}. As a result, we let $\Dot{e}$ denote the point where $K_3(e)$ attains its minimum in the interval $[0, e_{G_1}]$, and let $\hat{e}$ denote the point where $K_3(e)$ attains its minimum in the interval $[e_{G_2}, +\infty)$. 

However, for the interval $e\in(e_{G_1},e_{G_2})$, the monotonicity or convexity properties of $K_3(e)$ are hard to infer. Next, we will analyze the minimum value by assuming that whether there is an interior minimum point in $(e_{G_1},e_{G_2})$. Based on Assumption \ref{ass:uniqueTVaR}, the discussion is divided into the following three cases:
        \begin{enumerate}
            \item[(a)] If $e_{G_1}={\arg\min}_{e\in[e_{G_1},e_{G_2}]}K_3(e)$, then on the interval $[0,e_{G_2})$, the minimum is attained at $\Dot{e}$, and we need to compare the values of $K_3(\Dot{e})$ and $K_3(\hat{e})$:
            \begin{itemize}
                \item if $K_3(\Dot{e})\leq K_3(\hat{e})$, then the global minimum is attained at $\Dot{e}$;
                \item if $K_3(\Dot{e})> K_3(\hat{e})$, then the global minimum is attained at $\hat{e}$.
            \end{itemize}
            \item[(b)] If $e_{G_2}={\arg\min}_{e\in[e_{G_1},e_{G_2}]}K_4(e)$, then on the interval $(e_{G_1},+\infty)$, the minimum is attained at $\hat{e}$, and we need to compare the values of $K_3(\Dot{e})$ and $K_3(\hat{e})$:
            \begin{itemize}
                \item if $K_3(\Dot{e})\leq K_3(\hat{e})$, then the global minimum is attained at $\Dot{e}$;
                \item if $K_3(\Dot{e})> K_3(\hat{e})$, then the global minimum is attained at $\hat{e}$.
            \end{itemize}
            \item[(c)] If there exists an interior point $\tilde{e}$ such that  $\tilde{e}={\arg\min}_{e\in(e_{G_1},e_{G_2})}K_3(e)$, then we need to compare $K_3(\Dot{e})$, $K_3(\tilde{e})$ and $K_3(\hat{e})$:
            \begin{itemize}
                \item if $K_3(\Dot{e})\leq\min\{K_3(\tilde{e}), K_3(\hat{e})\}$, then the global minimum is attained at $\Dot{e}$;
                \item if $K_3(\tilde{e})\leq\min\{K_3(\Dot{e}), K_3(\hat{e})\}$, then the global minimum is attained at $\tilde{e}$;
                \item if $K_3(\hat{e})\leq\min\{K_3(\Dot{e}), K_3(\tilde{e})\}$, then the global minimum is attained at $\hat{e}$.
            \end{itemize}
        \end{enumerate}
To sum up, the proof for this case is finished.
\hfill $\blacksquare$

\end{document}